\documentclass[onecolumn,structabstract]{aa}
\usepackage[latin1]{inputenc}
\usepackage[pdftex,final]{graphicx}
\usepackage{amsmath}
\usepackage{txfonts}
\usepackage[comma,authoryear]{natbib}
\usepackage{hyperref}
\usepackage{xcolor}
\newcommand{\ie}{\textit{i.e.}}
\newcommand{\eg}{\textit{e.g.}}

\newcommand{\OmegaK}{\Omega_\mathrm{K}}
\newcommand{\OmegaC}{\Omega_\mathrm{C}}

\newcommand{\dpart}[2]{\frac{\partial #1}{\partial #2}}

\newcommand{\dphi}{\partial_{\phi}}

\newcommand{\vect}{\vec}
\newcommand{\lapl}{\Delta}

\newcommand{\grad}{\vect{\nabla}}
\renewcommand{\div}{\vect{\nabla}\cdot}
\renewcommand{\d}{\partial}

\renewcommand{\l}{\ell}

\newcommand{\az}{\vect{a}_{\zeta}}
\newcommand{\at}{\vect{a}_{\theta}}
\newcommand{\ap}{\vect{a}_{\phi}}

\newcommand{\eo}{\vect{e}_{\mathrm{obs.}}}
\newcommand{\ez}{\vect{e}_z}
\newcommand{\er}{\vect{e}_r}
\newcommand{\et}{\vect{e}_{\theta}}
\newcommand{\ep}{\vect{e}_{\phi}}
\newcommand{\es}{\vect{e}_s}
\newcommand{\Ez}{\vect{E}_{\zeta}}
\newcommand{\Et}{\vect{E}_{\theta}}
\newcommand{\Ep}{\vect{E}_{\phi}}
\newcommand{\n}{\vect{n}}
\newcommand{\xir}{\xi_r}
\newcommand{\xit}{\xi_{\theta}}
\newcommand{\xip}{\xi_{\phi}}
\newcommand{\xicz}{\xi^{\zeta}}
\newcommand{\xict}{\xi^{\theta}}
\newcommand{\xicp}{\xi^{\phi}}
\newcommand{\txicz}{\tilde{\xi}^{\zeta}}
\newcommand{\txict}{\tilde{\xi}^{\theta}}
\newcommand{\txicp}{\tilde{\xi}^{\phi}}
\newcommand{\dr}{\partial_r}
\newcommand{\dz}{\partial_{\zeta}}
\newcommand{\dt}{\partial_{\theta}}
\newcommand{\dzz}{\partial^2_{\zeta\zeta}}
\newcommand{\dzt}{\partial^2_{\zeta\theta}}
\newcommand{\dtt}{\partial^2_{\theta\theta}}
\newcommand{\dph}{\partial_{\phi}}
\newcommand{\dpp}{\partial^2_{\phi\phi}}
\newcommand{\cz}{c_{\zeta}}
\newcommand{\rz}{r_{\zeta}}
\newcommand{\rt}{r_{\theta}}
\newcommand{\rzz}{r_{\zeta\zeta}}
\newcommand{\rzt}{r_{\zeta\theta}}
\newcommand{\rtt}{r_{\theta\theta}}

\newcommand{\sini}{\sin i}
\newcommand{\cosi}{\cos i}
\newcommand{\sint}{\sin\theta}
\newcommand{\cost}{\cos\theta}
\newcommand{\cott}{\cot\theta}
\newcommand{\sinp}{\sin\phi}
\newcommand{\cosp}{\cos\phi}

\newcommand{\geff}{g_{\mathrm{eff}}}
\newcommand{\Teff}{T_{\mathrm{eff}}}
\newcommand{\hgeff}{\hat{g}_{\mathrm{eff}}}
\newcommand{\hTeff}{\hat{T}_{\mathrm{eff}}}
\newcommand{\vgeff}{\vect{g}_{\mathrm{eff}}}

\newcommand{\vlp}{\left[\omega + m\Omega\right]}
\newcommand{\ds}{\partial_s}

\newcommand{\dP}{\frac{\delta p}{P_0}}
\newcommand{\drho}{\frac{\delta \rho}{\rho_0}}

\newcommand{\Rp}{R_{\mathrm{p}}}
\newcommand{\Req}{R_{\mathrm{eq}}}
\newcommand{\Gz}{\mathcal{G}_{\zeta}}
\newcommand{\Gt}{\mathcal{G}_{\theta}}

\newcommand{\Ylm}{Y_{\ell}^m}

\begin{document}

\title{Mode visibilities in rapidly rotating stars}
\author{D. R. Reese\inst{1,2,3}
        \and
        V. Prat\inst{4,5}
        \and
        C. Barban\inst{2}
        \and
        C. van~'t Veer-Menneret\inst{6}
        \and
        K. B. MacGregor\inst{7}
       }

\institute{Institut d'Astrophysique et Géophysique de l'Université de Liège,
           Allée du 6 Août 17, 4000 Liège, Belgium \\
           \email{daniel.reese@ulg.ac.be}
           \and
           LESIA, Observatoire de Paris, CNRS, UPMC Univ. Paris 06,
            Univ. Paris-Diderot, 5 place Jules Janssen, 92195 Meudon, France
           \and
           Kavli Institute for Theoretical Physics, Kohn Hall, University of California, Santa Barbara, CA 93106, USA
           \and
           Universt{\'e} de Toulouse, UPS-OMP, IRAP, Toulouse, France
           \and
           CNRS, IRAP, 14 avenue Edouard Belin, 31400 Toulouse, France
           \and
           GEPI, Observatoire de Paris-Meudon, CNRS, Université Paris Diderot, 92125, Meudon Cedex, France
           \and
           High Altitude Observatory, National Center for Atmospheric Research, Boulder, CO 80307, USA
          }
\date{}

\abstract
{Mode identification is a crucial step to comparing observed frequencies with
theoretical ones.  However, it has proven to be particularly difficult in
rapidly rotating stars.  An important reason for this is the lack of simple
frequency patterns such as those present in solar-type pulsators.  This problem
is further aggravated in $\delta$ Scuti stars by their particularly rich
frequency spectra.}
{As a first step to obtaining further observational constraints towards mode
identification in rapid rotators, we aim to accurately calculate mode
visibilities and amplitude ratios while fully taking into account the effects of
rotation.}
{We derive the relevant equations for calculating mode visibilities in different
photometric bands while fully taking into account the geometric distortion from
both the centrifugal deformation and the pulsation modes, the variations in
effective gravity, and an approximate treatment of the temperature variations,
given the adiabatic nature of the pulsation modes.  These equations are then
applied to 2D oscillation modes, calculated using the TOP code (Two-dimension
Oscillation Program), in fully distorted 2D models based on the SCF
(Self-Consistent Field) method.  The specific intensities come from a grid of
Kurucz atmospheres, thereby taking into account limb and gravity darkening.}
{We obtain mode visibilities and amplitude ratios for $2\,M_{\odot}$ models with
rotation rates ranging from $0$ to $80\%$ of the critical rotation rate.  Based
on these calculations, we confirm a number of results from earlier studies, such
as the increased visibility of numerous chaotic modes at sufficient rotation
rates, the simpler frequency spectra with dominant island modes for pole-on
configurations, or the dependence of amplitude ratios on inclination and
azimuthal order in rotating stars.  In addition, we explain how the geometric
shape of the star leads to a smaller contrast between pole-on and equator-on
visibilities of equatorially-focused island modes.  We also show that modes
with similar $(\l,\,|m|)$ values frequently have similar amplitude ratios, even
in the most rapidly rotating models.}
{}

\keywords{stars: oscillations (including pulsations) -- stars: rotation}

\maketitle

\section{Introduction}

The space missions CoRoT \citep{Baglin2009, Auvergne2009} and Kepler
\citep{Borucki2009} are revealing very rich pulsational spectra in rapidly
rotating $\delta$ Scuti stars.  For instance, several hundred individual
frequencies have been found in HD 50844 and HD 181555, observed by CoRoT 
(\citealt{Poretti2009}, Michel, private communication), and V2367 Cyg, observed
by Kepler \citep{Balona2012}. It is becoming increasingly clear that
interpreting these spectra will not be a straightforward task and that theory is
lagging behind observations.  A crucial first step in interpreting this data is
correctly identifying the pulsation modes, \ie\  finding the correct
correspondence between theoretically calculated modes and observed pulsations. 
Recently, \citet{Reese2009b} proposed a way to identify acoustic pulsation modes
in rapidly rotating stars based on an asymptotic formula which describes the
frequencies of low degree modes \citep[see][and references therein]{Pasek2012}. 
Nonetheless this method runs into trouble if chaotic modes are present in the
pulsation spectra, which is expected based on the visibility calculations in
\citet{Lignieres2009}. Furthermore, the pulsation modes in $\delta$ Scuti stars
tend to be of low radial order, and may therefore be too far from the asymptotic
regime.  \citet{Lignieres2010} have worked on using the cross-correlation of
pulsation spectra.  Although it doesn't yield individual mode identifications,
it may provide a way of obtaining the rotation rate and/or the large frequency
separation, and explaining recent observations of recurring frequency spacings in
rapid rotators \citep{GarciaHernandez2009, Mantegazza2012}.  Nonetheless, the
need remains for methods capable of identifying individual pulsation modes.

Two particularly promising methods for identifying pulsation modes are
multi-colour photometric and spectroscopic mode identification.  The first
approach consists in measuring the amplitudes and phases of a given pulsation
mode in different photometric bands, calculating the ratios of the different
amplitudes and/or the phase differences, and comparing these to theoretical
predictions.  The second approach exploits the Doppler shifts caused by the
velocity field from the pulsation mode and how it affects observed absorption
lines.  These methods have been successfully applied to slowly rotating
stars \citep[\eg\ ][]{DeRidder2004,Zima2006,Briquet2007}, but more work is
needed before they are applied to rapid rotators.  In the present paper, we will
focus on mode visibilities in different photometric bands as a first step to
multi-colour photometric mode identification, and postpone spectroscopic mode
identification to a later paper.

Few studies have dealt with multi-colour photometric mode signatures in rapidly
rotating stars, and those that do generally approximate the effects of rotation
on the pulsation modes.  For instance, \citet{Daszynska_Daszkiewicz2002,
Daszynska_Daszkiewicz2007} and \citet{Townsend2003b} used either the
perturbative approach or the traditional approximation to calculate their
pulsation modes, and in some cases included the effects of avoided crossings.
Their calculations of mode visibilities included stellar surface distortion from
the pulsation modes and the Lagrangian perturbations to both the effective
temperature and gravity.  An interesting result from these studies is that
contrarily to non-rotating stars, amplitude ratios depend both on $m$, the
azimuthal order of the pulsation mode, and $i$, the inclination of the star.
More recently, \citet{Lignieres2006} and \citet{Lignieres2009} calculated
geometrical disk-integration factors of acoustic modes in deformed polytropic
models by integrating the temperature fluctuations over the visible disk.  The
effects of rotation were fully taken into account in the pulsation modes, thanks
to the 2D numerical approach, but non-adiabatic effects were neglected, thereby
making the fluctuations of the effective temperature inaccessible.  Furthermore,
gravity and limb darkening, the Lagrangian perturbations to the effective
gravity, and surface distortion caused by the modes were not taken into account.
Nonetheless, important first results were obtained through these articles,
namely, that chaotic modes are more visible than their non-rotating counterparts
due to irregular latitudinal node placement and may thus be detected, island
modes are the most visible modes in a pole-on configuration, leading to a
regular frequency pattern \citep{Lignieres2009}, and signatures of the large
frequency separation and/or rotation rate can show up in the autocorrelation
function of the frequency spectrum \citep{Lignieres2010}.

In order to obtain more realistic multi-colour photometric mode visibilities in
rapidly rotating stars, we derive a new set of equations which take into account
the Lagrangian variations to the effective temperature and gravity, as well as
the surface distortions induced both by the centrifugal deformation of the
equilibrium model and by the pulsation modes.  These are applied to adiabatic
acoustic modes calculated by the Two-dimension Oscillation Program
\citep[TOP,][]{Reese2006,Reese2009a} using rapidly rotating zero-age
main-sequence (ZAMS) models based on the Self-Consistent Field (SCF) method
\citep{Jackson2005, MacGregor2007}. The emergent intensities are calculated from
Kurucz atmospheres, taking into account the latitudinal dependence of the
equilibrium effective temperature and gravity, thereby including gravity and
limb darkening.  The main weakness in the present study is the adiabatic
approximation, which makes the Lagrangian fluctuations of the effective
temperature inaccessible.  As was previously done in \citet{Lignieres2009},
we approximate these by the Lagrangian temperature variations.  The
following section describes the pulsation calculations, with an emphasis on the
improvements and differences with the calculations done in \citet{Reese2009a}. 
This section is followed by a derivation of the relevant equations for
calculating mode visibilities in rapidly rotating stars.
Section~\ref{sect:photo_CoRoT} then describes various effects of rotation on
visibilities in a single band -- the CoRoT photometric band.  This is followed
by Sect.~\ref{sect:photo_multi} which deals with amplitude ratios in the Geneva
photometric system.  The paper ends with a short conclusion.

\section{Pulsation calculations}

In what follows, we review the methods used for obtaining the models and
associated pulsations that serve as inputs to the visibility calculations. 
These closely follow the approach used in \citet{Reese2009a} but include a
number of improvements as described below.

\subsection{Equilibrium models}
\label{sect:equilibrium_models}

The equilibrium models are calculated via the SCF method \citep{Jackson2005,
MacGregor2007}.  This method is an iterative procedure which alternates between
solving Poisson's equation and the equations of mass, momentum and energy
conservation before converging onto a 2D centrifugally deformed stellar model.
These models are chemically homogeneous ZAMS models with a cylindrical rotation
profile, although throughout the rest of the article, we will work with
uniformly rotating SCF models, even if the formulas in the visibility
calculations are established for general (non-cylindrical) rotation profiles. 
Given the rotation profile, the structure of the model is barotropic, \ie\ all
thermodynamic quantities remain constant on isopotentials, which are calculated
from the sum of the gravitational and centrifugal potentials.  Finally, we wish
to make the distinction between the critical rotation rate, $\OmegaC$, and the
Keplerian break-up rotation rate, $\OmegaK$:
\begin{equation}
\OmegaC = \sqrt{\frac{g_{\mathrm{eq}}}{\Req}}, \qquad
\OmegaK = \sqrt{\frac{GM}{\Req^3}}.
\end{equation}
Although very similar, the former uses the true gravity (excluding the
centrifugal force) at the equator, $g_{\mathrm{eq}}$, based on the actual
distribution of matter, to calculate the break-up rotation rate, whereas the
latter uses its Keplerian approximation, $\frac{GM}{\Req^2}$, which amounts to
assuming spherical symmetry for the distribution of matter.  As pointed out in
\citet{Roxburgh2004}, the Keplerian approximation slightly underestimates the
true gravity, so that $\OmegaC > \OmegaK$.  Table~\ref{tab:Omega} gives the
relative differences between these two quantities for different rotation rates,
calculated two different ways.  The first method is based on global quantities
provided with the models, whereas the second involves recalculating the
gravitational potential from the density distribution and using this to
calculate the equatorial gravity.  A comparison of columns two and three, and
five and six gives an idea of the uncertainty on these values.  We also note
that the theoretical value of $\left(\OmegaC-\OmegaK\right)/\OmegaK$ at $\OmegaC = 0.00$
is $0$, since the star is spherically symmetric and the Keplerian approximation
is exact.

\begin{table}[htbp]
\begin{center}
\caption{Relative differences, $(\OmegaC-\OmegaK)/\OmegaK$, for selected
rotation rates, calculated with two different methods (see text for
details).\label{tab:Omega}}
\begin{tabular}{ccccccc}
\hline
\hline
$\OmegaC$ & \textbf{Method 1} & \textbf{Method 2} & $\qquad$ &
$\OmegaC$ & \textbf{Method 1} & \textbf{Method 2} \\
\hline
$0.00$ &  -- & $1.1\times 10^{-4}$ & &
$0.50$ & $5.0\times 10^{-4}$ & $4.7\times 10^{-4}$ \\
$0.10$ & $9.5\times 10^{-5}$ & $4.9\times 10^{-5}$ & &
$0.60$ & $5.7\times 10^{-4}$ & $5.5\times 10^{-4}$ \\
$0.20$ & $2.3\times 10^{-4}$ & $2.1\times 10^{-4}$ & &
$0.70$ & $6.0\times 10^{-4}$ & $5.8\times 10^{-4}$ \\
$0.30$ & $3.0\times 10^{-4}$ & $2.7\times 10^{-4}$ & &
$0.80$ & $4.9\times 10^{-4}$ & $4.4\times 10^{-4}$ \\
$0.40$ & $3.1\times 10^{-4}$ & $2.6\times 10^{-4}$ & & & & \\
\hline
\end{tabular}
\end{center}
\end{table}

Before being used in the pulsation calculations, the models need to be
interpolated onto a new grid, and a number of supplementary equilibrium
quantities have to be derived, including a variety of geometric terms as well as
gradients of different equilibrium quantities.  Since \citet{Reese2009a}, a
number of improvements have been incorporated into these procedures.  For
instance, the stellar models are now interpolated onto a non-uniform radial grid
which becomes dense near the stellar surface. This allows the pulsation code to
correctly resolve the rapid spatial variations of acoustic modes near the
surface, resulting from the decrease in sound velocity.  Instead of
interpolating the density and pressure directly, their logarithm is
interpolated.  This leads to more accurate and consistent values near the
surface, where these quantities are several orders of magnitude smaller than in
the centre, and ensures they remain positive.  The effective gravity is
calculated via Poisson's equation.  This avoids taking the ratio of the pressure
gradient divided by the density, both of which are small quantities subject to
relatively large uncertainties.  Furthermore, the equipotentials, and hence the
geometric structure of the star, are recalculated using the solution from
Poisson's equation thus removing some numerical inaccuracies in the original
models.  The $\zeta$ derivative of equilibrium quantities is now correctly
calculated.  In \citet{Reese2009a}, the derivative was mistakenly calculated
with respect to $r$ rather than $\zeta$. Although this changed the quantitative
results, the qualitative conclusions from \citet{Reese2009a} remain unaltered. 
Finally, the $\Gamma_1$ profile is not derived from Eq.~(16) of
\citet{Jackson2005}, but rather from the equation of state,
which is based on the formula of \citet{Eggleton1973}.  Furthermore, the
quantity $\Gamma_2$, which intervenes in the visibility calculations described
below, is also calculated via the equation of state.

\subsection{Pulsation equations}
\label{sect:pulsation_equations}

A new set of variables is used in the pulsation equations:
\begin{equation}
\vect{\xi}, \qquad \dP, \qquad \drho, \qquad \Psi,
\label{eq:variables}
\end{equation}
where $\vect{\xi}$ is the Lagrangian displacement, $\delta p/P_0$ the Lagrangian
pressure perturbation divided by the equilibrium pressure profile,  $\delta
\rho/\rho_0$ the Lagrangian density perturbation divided by the equilibrium
density profile, and $\Psi$ the Eulerian perturbation to the gravitational
potential. Throughout this article, the subscript ``$0$'' denotes equilibrium
quantities.  We assume that the time and $\phi$ dependence of these variables
takes on the form $e^{i(\omega t + m\phi)}$, where $m$ is the azimuthal order. 
As such, we use what could be called the ``retrograde convention'', \ie\ modes
with positive azimuthal orders, $m$, are retrograde. 

Based on these variables, the continuity equation becomes
\begin{equation}
0 = \drho + \div \vect{\xi},
\label{eq:continuity}
\end{equation}
and Poisson's equation is
\begin{equation}
0 = \lapl \Psi - 4\pi G \left(\rho_0\drho - \vect{\xi}\cdot\grad\rho_0\right),
\label{eq:Poisson}
\end{equation}
where $G$ is the gravitational constant.  Euler's equation takes some more manipulations
\citep[see, for example, Eq.~A.3 of][]{Reese2009a}:
\begin{eqnarray}
-\vlp^2 \vect{\xi} &+& 2i \vlp \vect{\Omega} \times \vect{\xi}
      + \xi_s s \d_s \left(\Omega^2\right) \vect{e}_s 
  =  -\frac{\grad p}{\rho_0} + \frac{\rho \vect{g}_{\mathrm{eff}}}{\rho_0} - \grad \Psi \nonumber \\
 &=& -\frac{1}{\rho_0} \grad \left( P_0 \dP - \vect{\xi} \cdot \grad P_0\right)
     +\left(\rho_0 \drho - \vect{\xi} \cdot \grad \rho_0\right) \frac{\grad P_0}{\rho_0^2}
     -\grad \Psi \nonumber \\
 &=& -\frac{P_0}{\rho_0} \grad \left( \dP \right) 
     -\dP \frac{\grad P_0}{\rho_0}
     +\frac{\grad \left( \vect{\xi} \cdot \grad P_0 \right)}{\rho_0}
     +\drho\frac{\grad P_0}{\rho_0}
     -\frac{\left(\vect{\xi}\cdot\grad\rho_0\right)\grad P_0}{\rho_0^2}
     -\grad\Psi \nonumber \\
 &=& -\frac{P_0}{\rho_0} \grad \left( \dP \right) 
     +\left(\drho-\dP\right) \vgeff
     +\grad\left(\vect{\xi}\cdot\vgeff\right)
     -\grad\Psi
     +\left\{\frac{\left(\vect{\xi}\cdot\grad   P_0\right)\grad\rho_0 -
            \left(\vect{\xi}\cdot\grad\rho_0\right)\grad P_0}{\rho_0^2}\right\},
\label{eq:Euler}
\end{eqnarray}
where $\Omega$ is the rotation profile, $s$ the distance to the rotation axis,
and $\vgeff = -\grad \Psi_0 + s \Omega^2 \es =\grad P_0/\rho_0$ the effective
gravity.  The term in curly brackets cancels out because $\grad P_0$ is parallel
to $\grad \rho_0$ in a barotropic stellar structure.  Finally, the adiabatic
relation takes on the following very simple form:
\begin{equation}
\drho = \frac{1}{\Gamma_1} \dP.
\label{eq:adiabatic}
\end{equation}
This last equation is then used to eliminate $\delta \rho/\rho_0$ in favour of
$\delta p/P_0$ throughout the differential system and thus to reduce the size of
the problem compared to what is obtained in \citet{Reese2009a}.  Explicit
expressions for Eqs.~(\ref{eq:continuity}), (\ref{eq:Poisson})
and~(\ref{eq:Euler}), using spheroidal coordinates (see
Sect.~\ref{sect:spheroidal_geometry}), are given in
App.~\ref{sect:spheroidal_pulsation_equations}.

Besides reducing the computational cost, using the variables in
Eq.~(\ref{eq:variables}) allows us to obtain a much cleaner derivation of
$\delta p/P_0$, and hence $\delta T/T_0$, near the surface. Indeed,
if one were to calculate this quantity from the Eulerian pressure perturbation,
they would apply the following relation:
\begin{equation*}
\dP = \frac{p + \vect{\xi} \cdot \grad P_0}{P_0}.
\end{equation*}
Near the surface, this involves the sum of two nearly opposite terms, divided
by a small quantity, thus leading to poor numerical results.

\subsection{Non-dimensionalisation}

Contrarily to \citet{Reese2009a}, we non-dimensionalise the SCF models in a more
classical way.  The following reference quantities are used as units of length,
density and pressure:
\begin{equation}
\Req,\qquad \rho_{\mathrm{ref}} = \frac{M}{\Req^3},\qquad
P_{\mathrm{ref}} = \frac{GM^2}{\Req^4},
\end{equation}
where $\Req$ is the equatorial radius and $M$ the mass.  These lead
to the following time scale:
\begin{equation}
t_{\mathrm{ref}} = \left(\frac{\Req^3}{GM}\right)^{1/2}
= \frac{1}{\Omega_{\mathrm{K}}},
\end{equation}
where $\Omega_{\mathrm{K}}$ is the Keplerian break-up rotation rate.  Hence, the
non-dimensional frequencies are directly $\omega/\Omega_{\mathrm{K}}$. With this
non-dimensionalisation, the preceding pulsation equations remain unchanged
except for Poisson's equation, which becomes
\begin{equation}
0 = \lapl \Psi - 4\pi \left(\rho_0\drho - \vect{\xi}\cdot\grad\rho_0\right).
\end{equation}

\subsection{Spheroidal geometry}
\label{sect:spheroidal_geometry}

As was done in \citet{Lignieres2006} and \citet{Reese2006}, a surface-fitting
coordinate system, $(\zeta,\theta,\phi)$, based on \citet{Bonazzola1998}, is
introduced.  This system is related to the usual spherical coordinates,
$(r,\theta,\phi)$, via the relation
\begin{equation}
r(\zeta,\theta) = (1-\varepsilon)\zeta + \frac{5\zeta^3-3\zeta^5}{2}
                  \left(R_s(\theta) - 1 + \varepsilon\right),
\end{equation}
where $R_s(\theta)$ corresponds to the surface and $\varepsilon = 1 - \Rp/\Req$
is a measure of the oblateness, $\Rp$ being the polar radius, and $\zeta$ is
comprised between $0$ and $1$.  As can be seen, $\zeta=1$ corresponds to the
stellar surface.  A second domain, with a spherical outer boundary, is added
around the star so as to simplify the boundary condition on the gravitational
potential:
\begin{equation}
\label{eq:second_domain}
r(\zeta,\theta) = 2\varepsilon + (1-\varepsilon)\zeta
                + (2\zeta^3-9\zeta^2+12\zeta-4)\left(R_s(\theta) - 1 -\varepsilon\right),
\end{equation}
where $\zeta \in [1, 2]$. For $\zeta=1$, Eq.~(\ref{eq:second_domain}) coincides
with the stellar surface, whereas for $\zeta=2$, it yields a sphere of radius
$2$ (or $2\Req$ in dimensional form).  For conciseness, we will use the
subscripts ``$\zeta$'' and ``$\theta$'' to denote derivatives of $r$ with
respect to these variables.  For example, $\rz = \partial r/\partial \zeta$ and
$\rzt = \partial^2 r/\partial \zeta \partial \theta$.

\subsection{Boundary conditions}
\label{sect:boundary_conditions}

The pulsation equations are supplemented by a number of boundary equations.
Regularity of the solutions is imposed in the centre.  The perturbation to the
gravitational potential is made to go to zero at an infinite distance
from the star.  This condition is imposed by extending $\Psi$ into the second
domain and matching it to a vacuum potential on the second domain's outer
spherical boundary as described in \citet{Reese2006}.  When extending $\Psi$
into the second domain, both it and its gradient need to be kept continuous
across the \textit{perturbed} stellar surface.  This can be achieved by imposing
the continuity of the Lagrangian perturbation to the gravitational potential and
its gradient at $\zeta=1$:
\begin{equation}
\Psi^{\mathrm{int}} + \vect{\xi} \cdot \grad \Psi^{\mathrm{int}}_0 =
\Psi^{\mathrm{ext}} + \vect{\xi} \cdot \grad \Psi^{\mathrm{ext}}_0, \qquad
\grad \Psi^{\mathrm{int}} + \vect{\xi} \cdot \grad \left( \grad \Psi^{\mathrm{int}}_0 \right) =
\grad \Psi^{\mathrm{ext}} + \vect{\xi} \cdot \grad \left( \grad \Psi^{\mathrm{ext}}_0 \right),
\end{equation}
where the superscripts ``int'' and ``ext'' correspond to ``just below'' and
``just above'' the stellar surface, respectively.  Given that the gradient of
the equilibrium gravitational potential is continuous, the first condition simplifies
to
\begin{equation}
\label{eq:bc_psi}
\Psi^{\mathrm{int}} = \Psi^{\mathrm{ext}}.
\end{equation}
The second condition can be simplified, using Poisson's equation, applied to the
equilibrium gravitational potential:
\begin{equation}
\label{eq:bc_psi_grad}
\dz \Psi^{\mathrm{int}} = \dz \Psi^{\mathrm{ext}} - \frac{4\pi G \rho_0 \zeta^2 \rz}{r^2+\rt^2}\xicz, \qquad
\dt \Psi^{\mathrm{int}} = \dt \Psi^{\mathrm{ext}},\qquad
\dphi \Psi^{\mathrm{int}} = \dphi \Psi^{\mathrm{ext}}.
\end{equation}
Here, $\xicz$ is the radial component of the Lagrangian displacement, when
decomposed over the alternate basis $(\az,\,\at,\,\ap)$ introduced in 
\citet[][see also Eq.~(\ref{eq:alternate_basis})]{Reese2006}. The first equation
is different from what was applied in \citet{Reese2009a}, since it takes into
account the contribution from a non-zero surface density.  However, given its
low value, the resultant difference is quite negligible. The latter two
equations are implied by Eq.~(\ref{eq:bc_psi}), making it unnecessary to impose
them.

Finally, the usual mechanical boundary condition on the Lagrangian pressure
perturbation, $\mathbf{\delta p = 0}$, has been replaced by a slightly different
condition:
\begin{equation}
\vect{E}^{\zeta} \cdot \grad \left( \dP \right) = 0,
\label{eq:mechanical_bc}
\end{equation}
where $\vect{E}^{\zeta}$ is a vector perpendicular to the surface (see
Eq.~(\ref{eq:contravariant_basis})). The above, modified mechanical
condition corresponds to setting the vertical gradient of $\delta p/P_0$,
rather than its value, to zero.  We note that \citet{Pesnell1990} and
\citet{Dupret2002} applied a similar condition for spherically symmetric stars. 
In order to avoid having a boundary condition with a radial derivative in it, we
calculate the dot product between Euler's equation and $\vect{E}^{\zeta}$ and
cancel out the term corresponding to the vertical gradient of  $\delta p/P_0$. 
This leads to a complicated expression which is given in spheroidal coordinates
in App.~\ref{sect:mechanical_bc}.

\begin{figure}[htbp]
\begin{center}
\includegraphics[height=7cm]{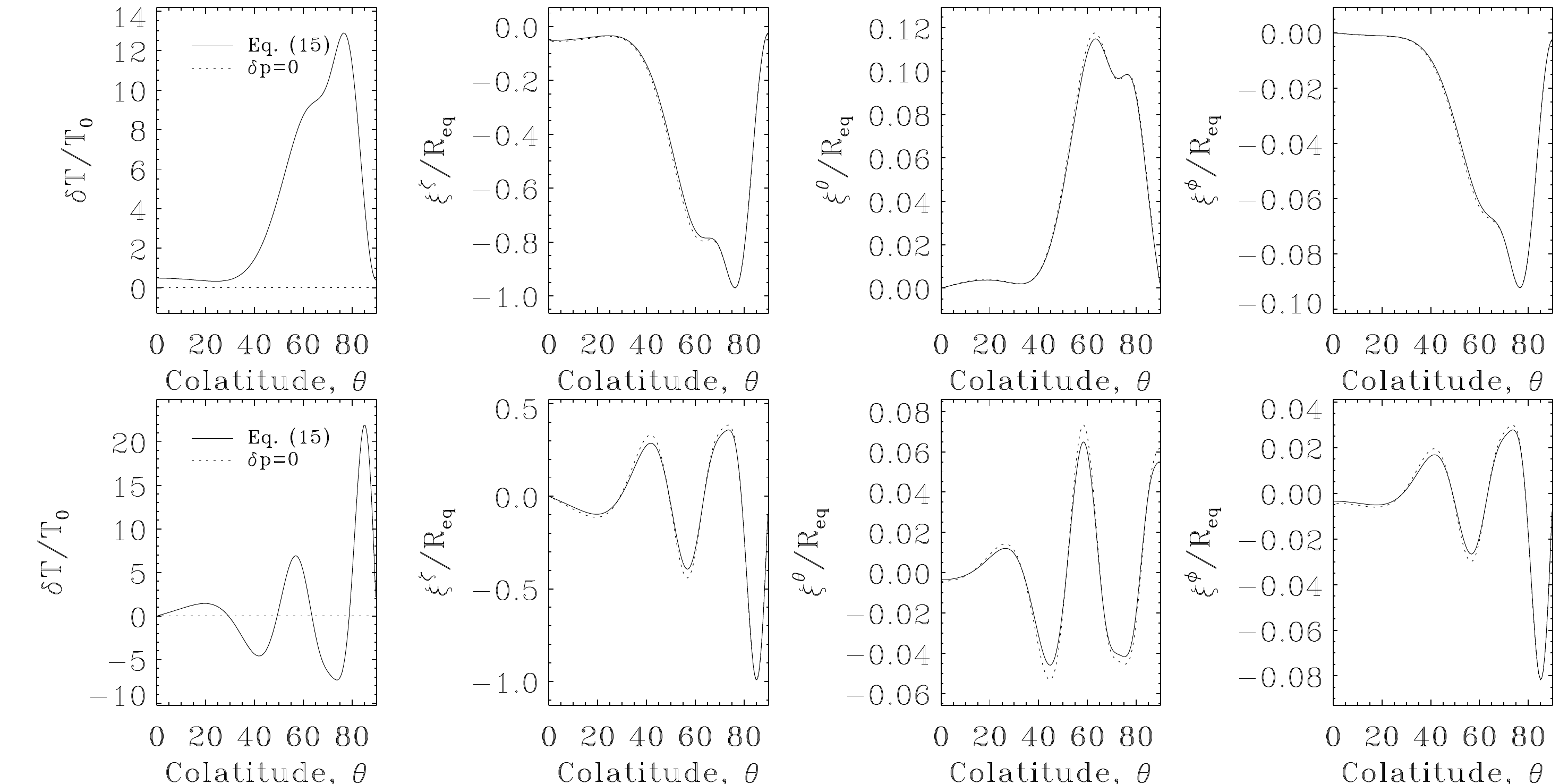} ~~
\includegraphics[height=7cm]{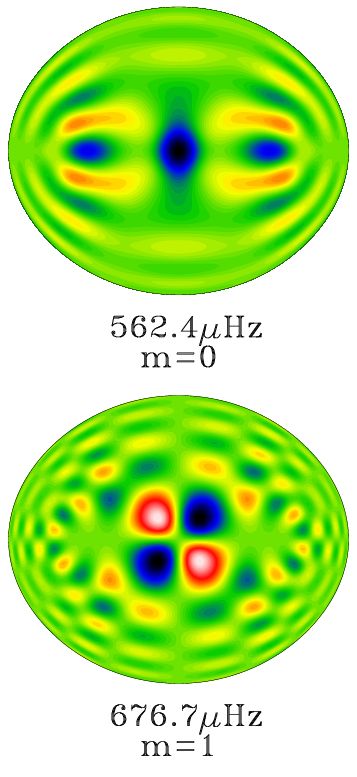}
\end{center}
\caption{Lagrangian temperature variations and Lagrangian displacement
at the stellar surface for two modes, using either $\delta p = 0$ or
Eq.~(\ref{eq:mechanical_bc}) as an external boundary condition.  The figures
to the right display  the meridional cross-section of the Eulerian pressure
perturbation of the two modes, divided by the square-root of the equilibrium
density.
\label{fig:bc}}
\end{figure}

The main purpose in using Eq.~(\ref{eq:mechanical_bc}) is to obtain
a non-zero value for $\delta T/T_0$, useful for visibility calculations as
described below.  Indeed, when combined with a non-zero surface pressure and the
adiabatic relation, the simpler condition, $\delta p = 0$, leads to $\delta
T/T_0 = 0$, as illustrated in Fig.~\ref{fig:bc} (left panels).  One may
then wonder if Eq.~(\ref{eq:mechanical_bc}) has an important effect on the
frequencies and on the displacement at the surface.  Numerically, it turns out
the frequencies vary little when using either boundary condition, at least in
the present study.  The middle panels of Fig.~\ref{fig:bc} also show that the
displacement is hardly affected.

\subsection{Numerical method}
\label{sect:numerical}

This set of equations and boundary conditions is projected onto the spherical
harmonic basis and discretised in the radial direction before being solved using
the code TOP \citep{Reese2006, Reese2009a}. Besides the improved way of treating
the equilibrium model, the present calculations also benefit from a new form of
finite differences.  This form achieves 4th order accuracy
for 1st order derivatives in spite of using windows with 4 rather 5 grid
points.  More importantly, this approach is robust to problems like mesh drift
and spurious solutions.

\section{Mode visibilities}
\label{sect:mode_visibilities}

At this point, we will switch to using the spherical vector basis
$(\er,\et,\ep)$ in which the polar or z-axis is lined up with the star's
rotation axis.  Furthermore, in order to make the equations more compact, we
will prefer the notation ``$r$'' to ``$R_s$'' when designating the stellar
surface, although it should be understood as $r(\zeta=1,\theta)$ in what
follows.  Similar implicit arguments also apply to other geometric terms such as
$\rz$ and $\rt$.  We introduce the unit vector $\eo$ which points \textit{from
the star to the observer}.  Furthermore, we will assume that the vector $\eo$
lies in the meridional plane $\phi=0$. Let $i$ be the inclination angle, \ie\
the angle between $\ez$ and $\eo$, where $\ez$ is lined up with the rotation
axis\footnote{As opposed to a non-rotating star, the inclination cannot be
arbitrarily set to $0$ to simplify the calculations.}.  An explicit expression
for $\eo$ in terms of the usual spherical basis is:
\begin{equation}
\eo = (\sini\sint\cosp+\cosi\cost)\er
    + (\sini\cost\cosp-\cosi\sint)\et
    - \sini\sinp\ep
\label{eq:eo}
\end{equation}

The radiated energy, received by an observing instrument from a non-pulsating stars, is:
\begin{equation}
E = \frac{1}{2\pi d^2}\iint_{\mathrm{Vis. Surf.}} I(\mu,\geff,\Teff) \eo \cdot \vect{\mathrm{d}S}
\label{eq:E}
\end{equation}
where $d$ is the distance to the star, $\mu$ the cosine of the angle between the
outward normal to the surface and $\eo$, $\geff$ and $\Teff$ the effective
gravity and temperature, and $I(\mu,\geff,\Teff)$ the specific radiation 
intensity, multiplied by the instrument's and/or filter's transmission curve,
and integrated over the wavelength spectrum,. As can be seen, the integral is
carried out over the visible surface.

In a pulsating star, this quantity is perturbed as follows:
\begin{equation}
\Delta E(t) = \frac{1}{2\pi d^2}\Re \left\{ \iint_{\Delta(\mathrm{Vis. Surf.})} I(\mu,\geff,\Teff) \eo \cdot \vect{\mathrm{d}S}
             + \iint_{\mathrm{Vis. Surf.}}  \delta I(\mu,\geff,\Teff,t) \eo \cdot \vect{\mathrm{d}S}
             + \iint_{\mathrm{Vis. Surf.}}  I(\mu,\geff,\Teff)  \eo \cdot \delta(\vect{\mathrm{d}S}) \right\}
\label{eq:E_perturbed}
\end{equation}
where $\delta$ denotes the Lagrangian perturbation, $\Re\{ \dots \}$ the real
part, and $\Delta(\mathrm{Vis. Surf.})$ the amount by which the visible
surface is modified due the modifications of the surface normal induced by the
oscillatory motions.  Furthermore, we assume a complex form for the
eigenfunctions, hence the reason for taking the real part of the above
expression.

The first term in Eq.~(\ref{eq:E_perturbed}) is proportional to the square of
the displacement and therefore neglected \citep[\eg\ ][]{Dziembowski1977}.

The Lagrangian perturbation to the specific intensity which intervenes in the
second term may be developed as follows:
\begin{equation}
\delta I = I \cdot \left(\dpart{\ln I}{\ln \Teff} \frac{\delta \Teff}{\Teff}
             + \dpart{\ln I}{\ln \geff} \frac{\delta \geff}{\geff} \right)
             + \dpart{I}{\mu} \delta \mu
\end{equation}
The partial derivatives of $I$ are obtained from model atmospheres and will be
dealt with in Sect.~\ref{sect:specific_intensities}.  We note that the above
expression is not exact as it neglects the slight Doppler shifts caused by
rotation and the oscillations.  Such shifts modify the position of the emerging
flux with respect to the instrument's and/or filter's transmission curve thereby
modifying $I$, but are expected to play a negligible role compared to other
effects.  We are therefore left with a number of geometrical terms to calculate
as well as the Lagrangian perturbation to the effective temperature and gravity.

\subsection{Geometrical terms}
In what follows, we will use the following expression for the displacement:
\begin{equation}
\vect{\xi} = \xir(\zeta,\theta,\phi,t) \er(\theta,\phi)
           + \xit(\zeta,\theta,\phi,t) \et(\theta,\phi)
           + \xip(\zeta,\theta,\phi,t) \ep(\phi)
\label{eq:spherical_components}
\end{equation}
where $(\er,\et,\ep)$ is the usual vector basis in spherical coordinates. 
\textit{Note}: we have used subscripts rather than superscripts, for the letters
$r$, $\theta$, and $\phi$, to distinguish these components from those given in
Eq.~(\ref{eq:alternate_components}).  In what follows, we will be using the
following relations:
\begin{equation}
\begin{array}{rclrclrcl}
\dt \er  &=& \et,             & \dt \et  &=& -\er,             & \dt \ep  &=& \vect{0}, \\
\dph \er &=& \sint \ep,\qquad & \dph \et &=& \cost \ep, \qquad & \dph \ep &=& -\sint \er - \cost \et
\end{array}
\end{equation}
We start by calculating a surface element on a rotating, non-pulsating star. 
This is given by the following expression:
\begin{equation}
\label{eq:dS}
\vect{\mathrm{d}S} = \left(\dt\vect{r}\times\dph\vect{r}\right) \mathrm{d}\theta \mathrm{d}\phi
          = \left(r^2\sint\er-r\rt\sint\et\right)\mathrm{d}\theta \mathrm{d}\phi.
\end{equation}
We then calculate the Lagrangian perturbation to a surface element:
\begin{eqnarray}
\label{eq:delta_dS}
\delta(\vect{\mathrm{d}S}) &=& \left(\dt\vect{\xi}\times\dph\vect{r} +
                      \dt\vect{r} \times\dph\vect{\xi}\right)\mathrm{d}\theta \mathrm{d}\phi \nonumber \\
                  &=& \left\{ \left(2r\sint\xir+r\cost\xit+r\sint\dt\xit+r\dph\xip\right)\er
                      +\left(-\rt\sint\xir-r\sint\dt\xir+(r\sint-\rt\cost)\xit-\rt\dph\xip\right)\et \right. \nonumber \\
                  & & \left. +\left(-r\dph\xir+\rt\dph\xit+(r\sint-\rt\cost)\xip\right)\ep\right\}\mathrm{d}\theta \mathrm{d}\phi.
\end{eqnarray}
The approach used to obtain Eqs.~(\ref{eq:dS}) and~(\ref{eq:delta_dS}) is
essentially the same as that of \citet{Buta1979} and \citet{Townsend1997}, but
the effects of horizontal Lagrangian displacements are also included in the
latter equation.  We then calculate $\mu$:
\begin{equation}
\mu = \eo \cdot \n
    = \frac{\left[r(\sini\sint\cosp+\cosi\cost) 
    - \rt(\sini\cost\cosp-\cosi\sint)\right]}
      {\left[r^2+\rt^2\right]^{1/2}},
\label{eq:mu}
\end{equation}
where $\n = \frac{\vect{\mathrm{d}S}}{\|\vect{\mathrm{d}S}\|}$.  The Lagrangian perturbation to
$\mu$ is calculated as follows:
\begin{eqnarray}
\delta\mu &=& \eo \cdot \delta \n
           = \eo \cdot \left\{ \frac{\delta \vect{\mathrm{d}S}}{\|\vect{\mathrm{d}S}\|} 
           - \frac{\vect{\mathrm{d}S}}{\|\vect{\mathrm{d}S}\|^2}\delta \|\vect{\mathrm{d}S}\| \right\}
           = \eo \cdot \left\{ \frac{\delta \vect{\mathrm{d}S}}{\|\vect{\mathrm{d}S}\|} 
           - \left( \n \cdot \frac{\delta \vect{\mathrm{d}S}}{\|\vect{\mathrm{d}S}\|} \right) \n \right \} \nonumber \\
          &=& \frac{\left[\rt(\sini\sint\cosp+\cosi\cost)
               +r(\sini\cost\cosp-\cosi\sint)\right]
              \cdot\left[\rt\xir-r\dt\xir+r\xit+\rt\dt\xit\right]}
              {\left[r^2+\rt^2\right]^{3/2}} \nonumber \\
          & & - \frac{\sini\sinp}{\sint} \cdot
              \frac{-r\dph\xir+\rt\dph\xit + (r\sint-\rt\cost)\xip}
              {r\left[r^2+\rt^2\right]^{1/2}},
\end{eqnarray}
where we have used the relation $\delta \|\vect{\mathrm{d}S}\| = \vect{\mathrm{d}S} \cdot \delta
\vect{\mathrm{d}S} / \|\vect{\mathrm{d}S}\| = \n \cdot \delta \vect{\mathrm{d}S}$. In the spherical limit,
the above expressions become
\begin{eqnarray}
\vect{\mathrm{d}S}         &=& \left(r^2\sint\er\right)\mathrm{d}\theta \mathrm{d}\phi, \\
\delta(\vect{\mathrm{d}S}) &=& \left\{ \left(2r\sint\xir+r\cost\xit+r\sint\dt\xit+r\dph\xip\right)\er
                      +\left(-r\sint\dt\xir+r\sint\xit\right)\et 
                      +\left(-r\dph\xir+r\sint\xip\right)\ep\right\}\mathrm{d}\theta \mathrm{d}\phi, \\
\mu               &=& \sini\sint\cosp+\cosi\cost, \\
\delta\mu         &=& \frac{\left[\sini\cost\cosp-\cosi\sint\right] \cdot \left[-\dt\xir+\xit\right]}{r}
                     -\frac{\sini\sinp}{\sint} \cdot \frac{-\dph\xir + \sint\xip}{r},
\end{eqnarray}
in full agreement with the expressions previously obtained by
\citet{Heynderickx1994} for a non-rotating star with $i=0$, provided one uses
the relations:
\begin{equation}
\xir = \delta r,\qquad \xit = r\delta\theta,\qquad \xip = r \sint\delta\phi.
\end{equation}

\subsection{Lagrangian perturbation to the effective gravity}
\label{sect:dgeff}

Although the pulsation equations are given for a cylindrical rotation profile,
we will relax this assumption in this section and allow for a general 2D
rotation profile when establishing an expression for the Lagrangian perturbation
to the effective gravity.  Before giving the Lagrangian perturbation to the
effective gravity, it is useful to recall various expressions for the unperturbed
surface effective gravity:
\begin{equation}
\vgeff = -\grad \Psi_0 + s\Omega^2 \es 
       = \frac{\grad P_0}{\rho_0}
       = \frac{\dz P_0}{\rho_0} \vect{E}^{\zeta}
       = - \geff \n.
\label{eq:geff}
\end{equation}
The minus sign in the last equation comes from the fact that gravity is pointed
inward, $\geff$ being the norm of $\vgeff$.  Furthermore, contrarily to what
happens in the stellar interior, the $\theta$ derivative of $P_0$ vanishes at
the surface since the surface is in pressure equilibrium.  Any of the
above expressions may be used to evaluate the surface effective gravity,
although $-\grad \Psi_0 + s\Omega^2\es$ provides the most accurate numerical
results.  It is also worth noting that the above expression neglects any
contributions from meridional circulation and viscous forces to the equilibrium
model \citep[\eg\ ][]{Rieutord2009}.

The relative Lagrangian perturbation to the effective gravity is then given by:
\begin{equation}
\frac{\delta \geff}{\geff} = \frac{\delta\|\vgeff\|}{\|\vgeff\|}
                           = -\frac{\n \cdot \delta\vgeff}{\geff},
\end{equation}
where we have used the simplification $\delta \|\vgeff\| = \vgeff \cdot \delta
\vgeff/\|\vgeff\| = -\n \cdot \delta \vgeff$. The quantity $\delta\vgeff$
represents the vectorial Lagrangian perturbation to the effective gravity.  It
includes the Lagrangian perturbation to the gradient of the gravitational
potential, and the acceleration of a particle \textit{tied to the surface},
resulting from the oscillatory motions:
\begin{equation}
\delta \vgeff = - \grad \Psi - \vect{\xi} \cdot \grad \left( \grad \Psi_0 \right)
                + (\omega+m\Omega)^2 \vect{\xi} 
                - 2i(\omega+m\Omega) \vect{\Omega}\times \vect{\xi}
                - \vect{\Omega} \times \left(\vect{\Omega} \times \vect{\xi} \right).
\label{eq:dvgeff_preliminary}
\end{equation}
Rather than working with the above expression, it is more useful to introduce
the equilibrium effective gravity by adding and subtracting $\vect{\xi} \cdot
\grad \left(s\Omega^2\es\right)$:
\begin{equation}
\delta \vgeff = - \grad \Psi + \vect{\xi} \cdot \grad \vgeff
                + (\omega+m\Omega)^2 \vect{\xi} 
                - 2i(\omega+m\Omega) \vect{\Omega}\times \vect{\xi}
                - \vect{\Omega} \times \left(\vect{\Omega} \times \vect{\xi} \right)
                - \vect{\xi} \cdot \grad \left(s\Omega^2\es\right).
\label{eq:dvgeff}
\end{equation}
After some simplifications based on Poisson's equation and a lengthy derivation
described in App.~\ref{sect:dgeff_appendix}, one obtains the following
explicit expression for a general rotation profile:
\begin{eqnarray}
\delta\geff &=& \frac{\left(r^2+\rt^2\right)^{1/2}\dz \Psi}{r\rz}
    -  \frac{\rt \dt \Psi}{r\left(r^2+\rt^2\right)^{1/2}}
    + \frac{r\xir - \rt\xit}{\left(r^2+\rt^2\right)^{1/2}}
       \left\{ 4\pi G \rho_0 - 2\Omega^2 
    -  \left[\frac{\sint\left(r\sint-\rt\cost\right)}{\rz} \dz \left(\Omega^2\right)
    +  \sint\cost\dt\left(\Omega^2\right) \right] \right\} \nonumber \\
   & & + \xir \left\{ \frac{\rt\dt\geff}{r^2+\rt^2}
    -  \frac{\left(2r-\rt\cott\right)\left(r^2+\rt^2\right)+r\rt^2-r^2\rtt}
      {\left(r^2+\rt^2\right)^2}\geff\right\}
    + \xit \left\{\frac{r\dt\geff}{r^2+\rt^2}
    -  \frac{r\rt\left(-2r^2-3\rt^2+r\rtt\right)+\left(r^2+\rt^2\right)\rt^2\cott}
      {r\left(r^2+\rt^2\right)^2}\geff\right\} \nonumber \\
   & & - (\omega+m\Omega)^2 \frac{r\xir-\rt\xit}{\left(r^2+\rt^2\right)^{1/2}} 
    + \left\{-2i(\omega+m\Omega)\Omega\xip 
    + \left[\frac{r\xir-\rt\xit}{\rz}\sint\dz\left(\Omega^2\right)
    + \xit\sint\dt\left(\Omega^2\right)\right]\right\}\frac{r\sint-\rt\cost}
      {\left(r^2+\rt^2\right)^{1/2}}.
\label{eq:dgeff}
\end{eqnarray}
It is worth noting that in the above expression, all of the terms involving a
$\zeta$ derivative of one of the perturbed quantities are divided by $\rz$.  The
remaining terms involve no $\zeta$ derivatives whatsoever.  This characteristic
is what one would expect for a quantity which is independent of the mapping.

If a cylindrical rotation profile is used instead, the terms in square brackets
simplify to the following expressions, respectively:
\begin{equation}
s\d_s\left(\Omega^2\right),\qquad
\left(\sint\xir + \cost\xit\right)s\d_s \left(\Omega^2\right).
\end{equation}

In the non-rotating limit, Eq.~(\ref{eq:dgeff}) reduces to:
\begin{equation}
\delta\geff = \dr \Psi + 4\pi G \rho_0\xir - \frac{2\xir}{r} \geff  - \omega^2\xir
\end{equation}
in full agreement with \citet{Dupret2002}.

\begin{figure}[htbp]
\includegraphics[width=\textwidth]{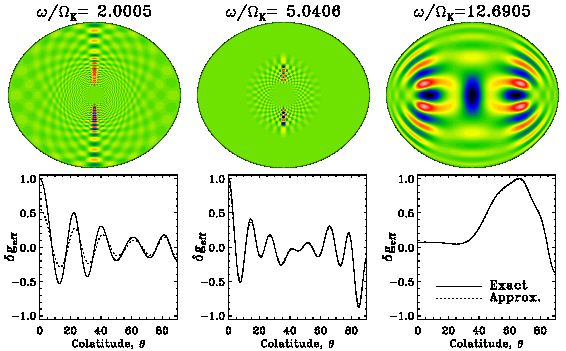}
\caption{Comparison between $\delta \geff$ and $\delta\geff^{\mathrm{approx.}}$
for three modes.  The upper row shows the meridional cross-section of
the three modes, whereas the lower row compares the two calculations
for the perturbation to the effective gravity at the stellar surface, between
one of the poles $(\theta=0^{\circ})$ and the equator $(\theta=90^{\circ})$. 
For the middle and right mode, the two curves nearly overlap, making it hard to
distinguish them.
\label{fig:dgeff}}
\end{figure}

Given the complexity of Eq.~(\ref{eq:dgeff}), it is interesting to see if it can
be approximated by a simpler expression.  We consider the following approximation:
\begin{equation}
\delta\geff^{\mathrm{approx.}} = -(\omega+m\Omega)^2 \vect{\xi} \cdot \n
    = - (\omega+m\Omega)^2 \frac{r\xir-\rt\xit}{\left(r^2+\rt^2\right)^{1/2}}.
\end{equation}
Figure~\ref{fig:dgeff} compares $\delta\geff$ and
$\delta\geff^{\mathrm{approx.}}$ for three different modes -- two gravito-inertial
modes and one p-mode.  As can be seen in this figure,
$\delta\geff^{\mathrm{approx.}}$ is a very good approximation for p- and
g-modes with a sufficiently high frequency.  It is only for low frequency
g-modes that the difference between the two expressions becomes non-negligible.

\subsection{Lagrangian perturbation to the temperature}

We then turn our attention to the relative Lagrangian perturbations of the
effective temperature, $\delta\Teff/\Teff$.  Given that the pulsation
calculations are done in the adiabatic approximation, this quantity is not
readily available.  We will therefore content ourselves with the approximation
$\delta\Teff/\Teff \approx \delta T/T_0$.  We note that this approximation is
different from the approximation $\delta \Teff \approx \delta T$ given that
$\Teff$ is latitude-dependant whereas $T_0$ is not in barotropic stellar
models.  As was pointed out in Sect.~\ref{sect:boundary_conditions}, a
modified version of the mechanical boundary condition was needed in order to
obtain a non-zero value for $\delta T$ at the surface.  One may then wonder if
other approximations may be more suitable, such as applying the boundary
condition $\delta p = 0$ high up in the stellar atmosphere and extracting
$\delta T$ at a deeper optical depth, or working with the Eulerian temperature
fluctuation instead.  The main difficulty with the first option is that the
stellar models used in the present study do not include an
atmosphere\footnote{The Kurucz atmospheres described in the following
section are only used to calculate the emergent intensities as a function of
$\Teff$ and $\geff$, and are not ``joined'' to the present models.}.  Hence, it
is not clear at what depth (which is likely to be latitude-dependant), one
should extract the Lagrangian temperature variations.  Using the Eulerian
temperature variations also has problems of its own.  Indeed, at the surface,
the Eulerian temperature variations are dominated by the the advection term,
thereby making them an order of magnitude larger than the Lagrangian variations,
and with the opposite sign for the most part.  Such variations do not seem the
most appropriate from a physical point of view, since the intensity variations
are more likely to result from local temperature variations physically present
in the fluid, \ie\ Lagrangian variations.  Hence, given the limitations of our
models and pulsation calculations, using the Lagrangian temperature variations
along with the modified mechanical boundary condition seems like the best
choice.  Nonetheless, according to \citet{Dupret2002} and \citet{Dupret2003},
adiabatic calculations of $\delta T/T_0$ are not reliable in the superficial
layers and consequently lead to a poor approximation of $\delta\Teff/\Teff$.  In
order to obtain $\delta \Teff/\Teff$ accurately, one would need to do a full
non-adiabatic calculations including a detailed treatment of the
atmosphere such as what is done in \citet{Dupret2002}, using 2D rotating models
in thermal equilibrium, such as what is being developed in the ESTER project
\citep{Rieutord2009, Espinosa2010}.

The quantity $\delta T/T_0$ can be deduced very simply from $\delta p/P_0$ via
the adiabatic relation:
\begin{equation}
\frac{\delta T}{T_0} = \frac{\Gamma_2 - 1}{\Gamma_2} \frac{\delta p}{P_0}
\end{equation}
where $\Gamma_2$ is the second adiabatic exponent, given by the expression
$\frac{\Gamma_2 - 1}{\Gamma_2} = \left(\dpart{\ln T}{\ln
p}\right)_{\mathrm{ad}}$.

\subsection{Specific intensities}
\label{sect:specific_intensities}

\begin{figure}[htbp]
\begin{center}
\begin{tabular}{cc}
\includegraphics[width=0.45\textwidth]{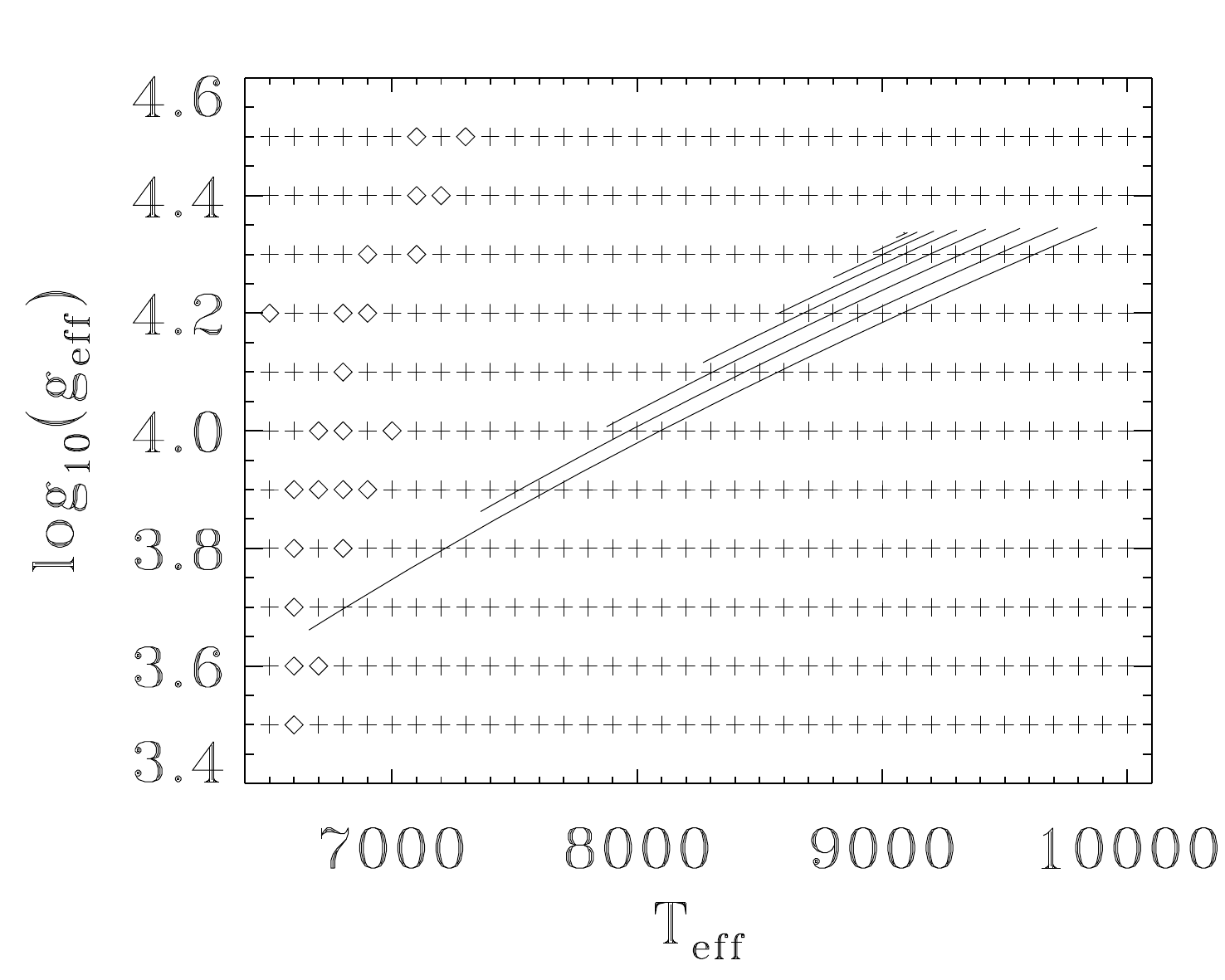} &
\includegraphics[width=0.45\textwidth]{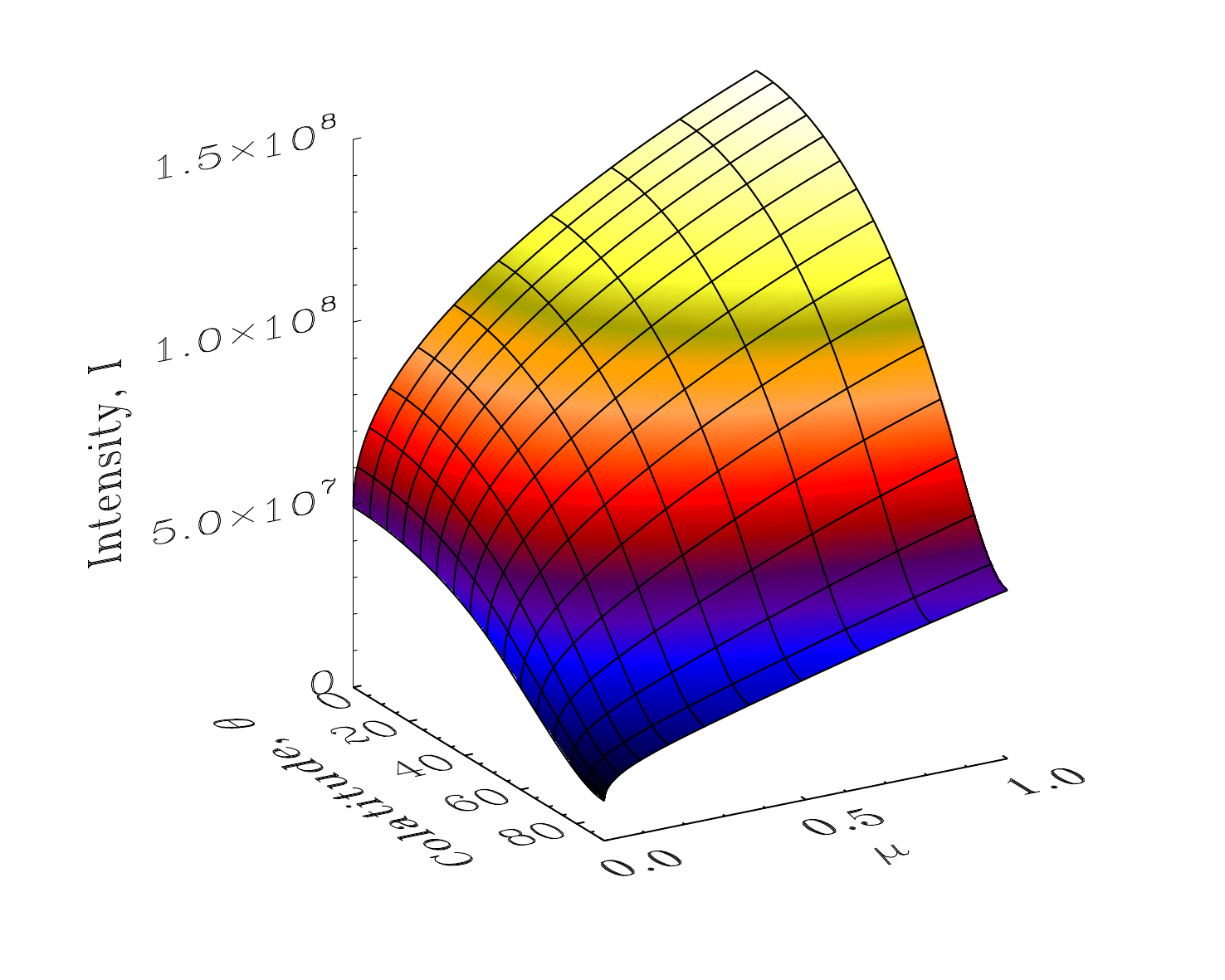} \\
\includegraphics[width=0.45\textwidth]{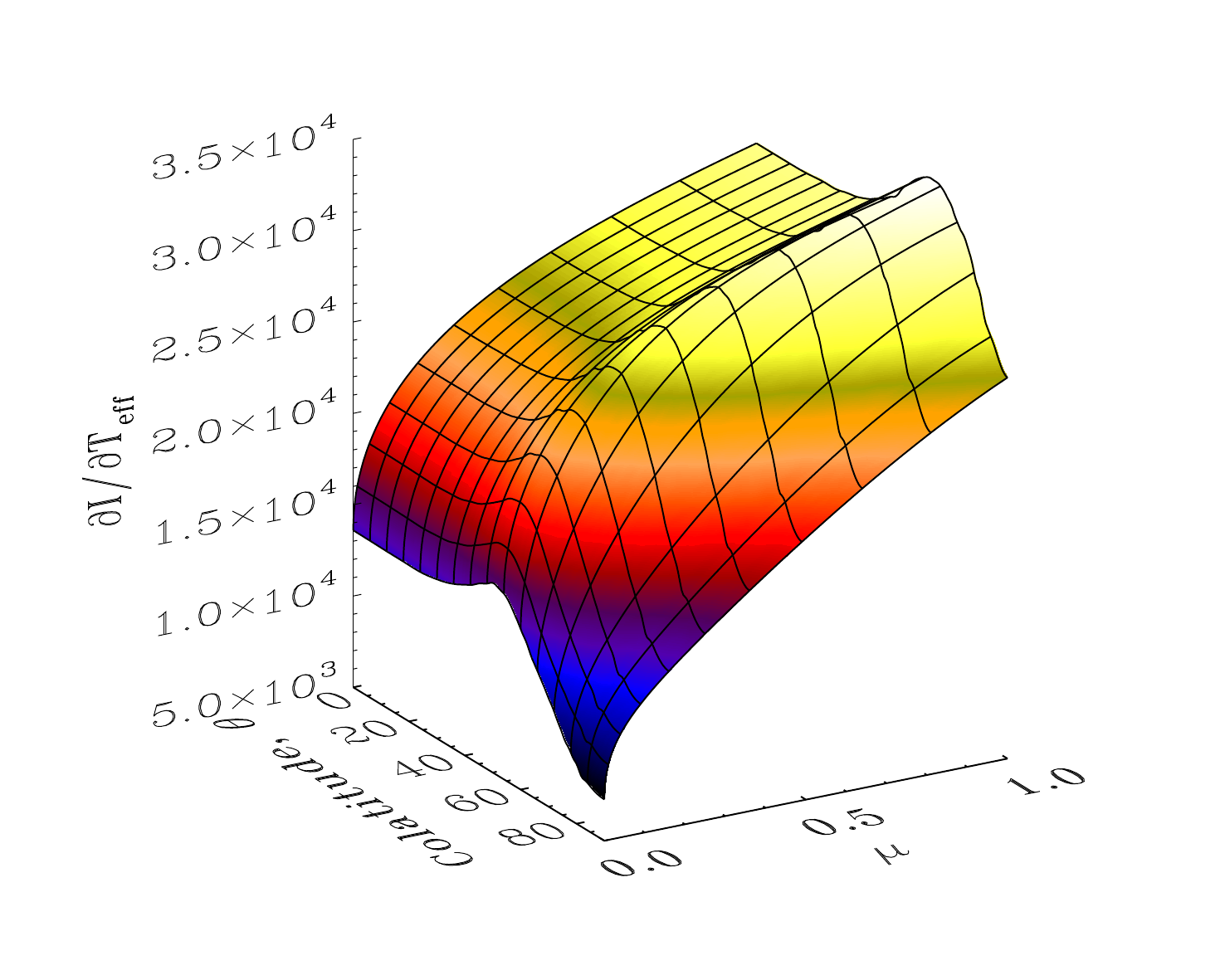} &
\includegraphics[width=0.45\textwidth]{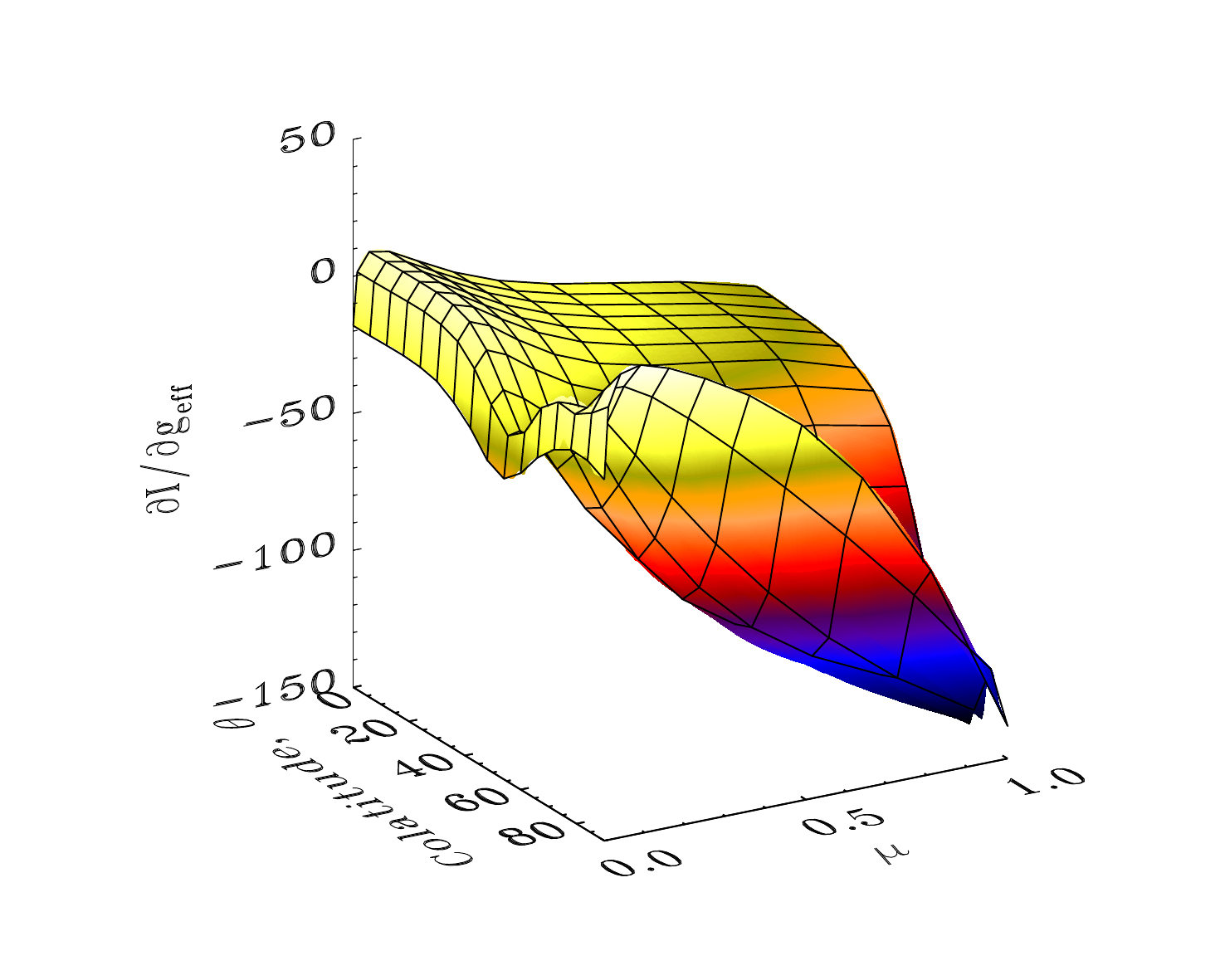}
\end{tabular}
\end{center}
\caption{\textbf{(Upper left panel)} Effective temperatures and gravities of the
Kurucz model atmospheres (symbols) and the $2\,M_{\odot}$ models (continuous
lines; the lower longer lines correspond to more rapid rotation). The pluses
$(+)$ represent initial Kurucz atmospheres whereas the diamonds $(\diamond)$
represent missing points which were interpolated from the neighbouring
atmosphere models. \textbf{(Remaining panels)} Specific intensities and its
partial derivatives for a $2\,M_{\odot}$, $0.8\,\Omega_\mathrm{c}$ model, as a
function of colatitude and $\mu$.
\label{fig:intensities}}
\end{figure}

Contrarily to the non-rotating case, the effective temperature and gravity of
the equilibrium model depend on the latitude.  Consequently, it is necessary to
find the appropriate set of specific intensities and their derivatives for each
latitude.  We therefore calculated a grid of Kurucz atmospheres with solar
composition which spans the relevant temperature and gravity ranges. We used the
ATLAS 9 code\footnote{See \url{http://kurucz.harvard.edu}} \citep{Kurucz1993}
in a modified form so as to include the convective prescription of
\citet{Canuto1996}, known as CGM (for more details see \citealt{Heiter2002} and
\citealt{Barban2003}).  Due to convergence problems, some of the grid points
were missing and had to be interpolated from neighbouring points.  The final
grid is illustrated in Fig.~\ref{fig:intensities} (upper left panel) as is
$\log_{10}\left(\geff\right)$ as a function of $\Teff$ for a set of
$2\,M_{\odot}$ models with rotation rates ranging from $0 \%$ to $80 \%$ of the
critical break-up rotation rate.

For each grid point, we then approximated the $\mu$ dependence through a Claret
type law \citep{Claret2000} using a least-squares fit.  This yielded a set of
five coefficients $\left(I(\mu=1),\,I(1)a_1,\,I(1)a_2,\, I(1)a_3 ,\,I(1)a_4
\right)$ from which it is possible to calculate $I$ and $\d I/\d \mu$ for any
$\mu$.  Using this approach rather than dealing directly with the original
values for $I$ yields better numerical results, both for interpolating $I$ and
especially for taking its $\mu$ derivative.

We then successively applied cubic spline interpolations, first as a function of
$\log_{10}\left(\geff\right)$, then as a function of $\Teff$.  This yielded
b-spline coefficients from which it is possible to deduce the Claret
coefficients and their derivatives for any value of $\log_{10} \left(\geff
\right)$ and $\Teff$ within the relevant range.  From this representation of
$I$, we then calculated the Claret coefficients for $I$, $\d I/\d \geff$ and $\d
I/\d \Teff$ as a function of latitude for each stellar model.  The remaining
three panels of Fig.~\ref{fig:intensities} display these quantities as a
function of $\mu$ and colatitude, $\theta$, for a $2\,M_{\odot}$ model at
$80\,\%$ of the critical rotation rate.  As is illustrated in these panels, both
limb and gravity darkening are taken into account.  The latitude-dependant
Claret coefficients were then projected onto the spherical harmonic basis,
truncated at 40 terms. This last step allows us to easily interpolate these
coefficients onto a denser latitudinal grid during the visibility calculations.

\subsection{Visibility integrals}

Once the above quantities are calculated over the stellar surface, the
visibilities may be evaluated numerically.  However, rather than using the time
dependant form given in Eq.~(\ref{eq:E_perturbed}), it is more useful to extract
the amplitude and phase of $\Delta E(t)$.  To do so, we cast
Eq.~(\ref{eq:E_perturbed}) into the following schematic form:
\begin{eqnarray}
\Delta E(t) &=& \Re \left\{ \iint_{\mathrm{Vis. Surf.}} \left[A(\theta,\phi) + i B(\theta,\phi)\right]
                e^{im\phi + i\omega t} \mathrm{d}\theta \mathrm{d}\phi \right\} \nonumber \\
            &=& \Re \left\{ \iint_{\mathrm{Vis. Surf.}} \left[
                 \left(A \cos (m\phi) - B \sin(m\phi)\right) + i \left( A \sin (m\phi) + B \cos(m\phi)\right) \right]
                 e^{i\omega t} \mathrm{d}\theta \mathrm{d}\phi \right\} \nonumber \\
            &=& \cos (\omega t) \underbrace{\iint_{\mathrm{Vis. Surf.}} \left(A \cos (m\phi) - B \sin(m\phi)\right) \mathrm{d}\theta \mathrm{d}\phi}_{\mathcal{A}}
             -  \sin (\omega t) \underbrace{\iint_{\mathrm{Vis. Surf.}} \left(A \sin (m\phi) + B \cos(m\phi)\right) \mathrm{d}\theta \mathrm{d}\phi}_{\mathcal{B}}
\end{eqnarray}
where $A(\theta,\phi)$ and $B(\theta,\phi)$ are real.  This leads to:
\begin{equation}
\Delta E(t) = \mathcal{C} \cos (\omega t + \varphi)
\end{equation}
where $\mathcal{C} = \sqrt{\mathcal{A}^2 + \mathcal{B}^2}$ is the amplitude and
$\varphi = \arctan \left(\mathcal{B}/\mathcal{A}\right)$ the phase.  Based on
symmetry considerations with respect to the variable $\phi$, the term
$\mathcal{B}$ cancels out in the adiabatic case, thereby leading to $\mathcal{C}
= |\mathcal{A}|$ and $\varphi = 0$ or $\pi$.  If, however, one tries to simulate
non-adiabatic effects by introducing a phase-lag in the temperature variations,
then the term $\mathcal{B}$ no longer cancels out.

\begin{figure}[htbp]
\includegraphics[width=\textwidth]{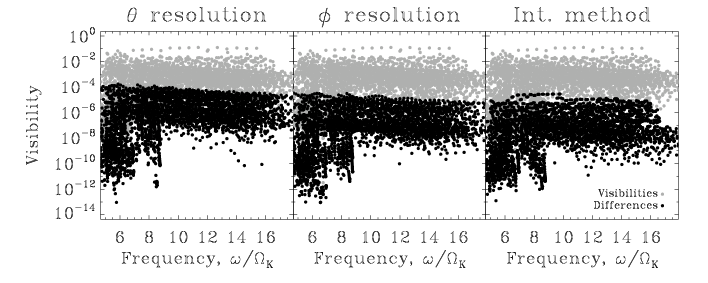}
\caption{Visibility calculations and differences for a set of $3197$ modes,
obtained using two different $\theta$ resolutions, $N_{\theta}=300$ and $2000$
\textit{(left panel)}, two different $\phi$ resolutions, $N_{\phi}=720$ and
$4000$ \textit{(middle panel)}, and two different integration methods in the
theta direction \textit{(right panel)}.  These modes were calculated in a
$2.0\,M_{\odot}$ model at $0.6\,\Omega_\mathrm{C}$, with an inclination angle
$i=10^{\circ}$.  If $i=80^{\circ}$ is used instead, the differences are reduced.
\label{fig:integration_errors}}
\end{figure}

The integrals $\mathcal{A}$ and $\mathcal{B}$ are evaluated numerically. The
surface is discretised in the $\theta$ direction, using a Gauss-Legendre
collocation grid with typically $N_{\theta} = 300$ points, and in the $\phi$
direction using a uniform grid with a typical resolution of $N_{\phi} = 720$. 
The integration weights in the $\theta$ direction are deduced from a Gauss
quadrature, whereas those in the $\phi$ direction are uniform.  
Figure~\ref{fig:integration_errors}, shows the differences which result from
increasing either $N_{\theta}$ (left panel) or $N_{\phi}$ (middle panel) in a
$2\,M_{\odot}$ stellar model, rotating at $60\,\%$ of the critical rotation
rate.  As can be seen, these differences are several orders of magnitude smaller
than the visibilities themselves.

At each point on the surface, the following condition is evaluated to determine
whether or not the surface element is facing the observer and should be included
in the integral:
\begin{equation}
\mu \geq 0
\label{eq:visible_surface}
\end{equation}
where $\mu$ is given in Eq.~(\ref{eq:mu}).  We note that this condition is only
valid if no obstacles are present between the surface element and the observer,
such as what could happen, for instance, in the more distorted SCF models with
concavities at the poles (due to highly differential rotation) and an
orientation which is not too far from equator-on. For such configurations, a more
general condition based on a z-buffer or ray-tracing approach would be required.

One possible concern is that using a Gauss quadrature implicitly assumes a
spectral decomposition of the function which is being integrated, which could
lead to a Gibbs phenomena near the cutoff between the visible and hidden side of
the star.  However, given that the Gibbs phenomena is oscillatory, one can
expect its integrated contribution do be small.  Nonetheless, we carried out a
test on the same model, using both Gauss quadrature and a trapezoidal
integration.  As can be seen in the right panel of
Fig.~\ref{fig:integration_errors}, the two approaches yielded very similar
results, showing that the Gibbs phenomena does not play an important role.

\subsection{Comparison with 1D case for non-rotating stars}

It is also important to check that in the non-rotating case, visibility
calculations based on the above 2D integrations agree with the results obtained
with a simpler 1D integration, as is commonly used in other works \citep[\eg\
][]{Heynderickx1994}.  In order to obtain the 1D integrals, one starts with a
given mode, which in this case will be proportional to a particular spherical
harmonic, $\Ylm(\theta,\phi)$, re-expresses this spherical harmonic in terms of a
new coordinate system with the $z'$-axis aligned with the line of sight via the
following decomposition \citep{Edmonds1960}:
\begin{equation}
\Ylm(\theta,\phi) = \sum_{m'=-\l}^{\l} d^{\l}_{mm'} (i) Y_{\l}^{m'}(\theta',\phi'),
\end{equation}
and calculates the integrated contribution from each of the
$Y_{\l}^{m'}(\theta',\phi')$.  In this new coordinate system, the visible
surface corresponds to a hemisphere.  Consequently, all of the non-axisymmetric
terms $m' \neq 0$ cancel out and only the term $d^{\l}_{m0} (i)
Y_{\l}^{0}(\theta',\phi')$ remains.  This leads to the following expression:
\begin{equation}
\Delta E(t) = \frac{r^2 d^{\l}_{m0}(i)}{d^2} \Re\left\{ \left[\frac{\delta \hTeff}{\Teff} \int_0^1 \dpart{I}{\ln\Teff} Y_0^{\l} \mu' \mathrm{d}\mu'
         + \frac{\delta \hgeff}{\geff} \int_0^1 \dpart{I}{\ln\geff} Y_0^{\l} \mu' \mathrm{d}\mu'
         + \frac{\hat{\xi}_r}{r} (2+\l)(1-\l) \int_0^1 I Y_0^{\l} \mu' \mathrm{d}\mu' \right] e^{i\omega t}\right\}
\label{eq:1D_integration}
\end{equation}
where
\begin{equation*}
d^{\l}_{m0}(i) = \varepsilon \sqrt{\frac{4\pi}{2\l+1}} \Ylm (\theta=i,\phi=0), \qquad
\mu' = \cos\theta', \qquad
\xi_r = \hat{\xi}_r \Ylm, \qquad
\delta \Teff = \left(\delta \hTeff\right) \Ylm, \qquad \mbox{and} \qquad
\delta \geff = \left(\delta \hgeff\right) \Ylm,
\end{equation*}
and $\varepsilon = \pm 1$ depending on the sign convention used in defining the
spherical harmonics, and the values of $\l$ and $m$. 
Figure~\ref{fig:1D_comparison} shows the differences between the two
approaches.  As can be seen, the results are very similar, thereby validating
the 2D integrations.

\subsection{Comparison with Lignières et al. (2006) and Lignières \& Georgeot (2009)}
\label{sect:comparision_Francois_me}

\begin{figure}[htbp]
\begin{tabular}{lr}
\includegraphics[width=0.47\textwidth]{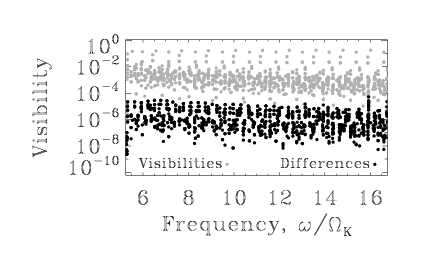} &
\includegraphics[width=0.47\textwidth]{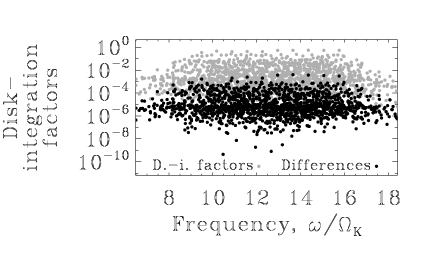} \\
\parbox[][4cm][t]{0.46\textwidth}{
\caption{Visibility calculations and differences for a set of $1223$ modes in a
non-rotating model, using 1D and 2D integrations.  The latitudinal resolution is
$N_{\theta} = 300$ for both approaches, and $N_{\phi} = 720$ points are used in
the azimuthal direction for the 2D integrations. The inclination angle is
$i=10^{\circ}$. \textit{Note:} because of mode degeneracy, there are much fewer
frequencies. \label{fig:1D_comparison}}} &
\parbox[][4cm][t]{0.46\textwidth}{
\caption{Simplified disk-integration factors and differences for a set of $1507$
modes in a model rotating at $0.8\,\Omega_{\mathrm{C}}$, using 2D integrations
and the harmonic projection method described in \citet{Lignieres2006} and
\citet{Lignieres2009}.  The resolutions are $N_{\theta} = 300$ for both
approaches, and $N_{\phi} = 720$ for the 2D integrations. The inclination angle
is $i=10^{\circ}$. \textit{Note:} due to symmetries in the simplified visibility
calculations, the visibilities of some modes cancel out and were therefore not
included in this plot (see text for details). 
\label{fig:comparison_Francois_me}}}
\end{tabular}
\end{figure}

Before going on to describe the results, we carry out one last test in which we
compare the 2D integration method described here with the harmonic projection
method used in \citet{Lignieres2006} and \citet{Lignieres2009}.  In order to
carry out the comparison, we calculate simplified disk-integration factors,
$D(i)$, by only integrating the temperature fluctuations over the visible
surface.  We avoid using the intensity fluctuations deduced from the Kurucz
atmospheres, as this would introduce the effects of limb darkening and a $\phi$
dependence of a different form than $e^{i m \phi}$, which is not currently
implemented in the method by \citet{Lignieres2009}.  Hence, the disk-integration
factors are given by:
\begin{equation}
D(i)\cos(\omega t + \psi) = \Re\left\{\frac{1}{\pi \Req^2 \left< \delta T \right>} \iint_{\mathrm{Vis. Surf.}} 
                            \delta T \eo \cdot \vect{\mathrm{d}S}\right\},
\qquad \mbox{where} \qquad
\left< \delta T \right> = \left( \frac{1}{S} \iint_{\mathrm{S}} \left|\delta T\right|^2 \mathrm{d}S \right)^{1/2},
\label{eq:visibility_francois}
\end{equation}
and where $S$ is the total surface and $\psi$ a suitably chosen phase.  The
disk-integration factors are normalised in the same way as is done in
\citet{Lignieres2006} and \citet{Lignieres2009}, which leads to $D(0) = 1$ if
$\delta T \equiv 1$.  Figure~\ref{fig:comparison_Francois_me} shows the
differences between the two approaches for a model rotating at
$0.8\,\Omega_{\mathrm{C}}$.  As can be seen, the differences are quite small
relative to the visibilities themselves.  At this point, it is helpful to
remember that the curve which delimits the visible surface, is approximated as
the intersection of a plane and the stellar surface in the harmonic projection
method.  One can expect such an approximation to become less good at high
rotation rates due to the strong centrifugal deformation, but as shown in
Fig.~\ref{fig:comparison_Francois_me} the effect remains small, in full
agreement with the results presented in \citet{Lignieres2009}, based on a Roche
model.

Given the simplified form given in Eq.~(\ref{eq:visibility_francois}), it turns
out that modes with $|m| \geq 2$ and such that $(-1)^{m+1} \delta \hat{T}(\pi -
\theta) = \delta \hat{T} (\theta)$, \ie\ even modes when $m$ is odd and
vice-versa, have zero disk-integration factors, regardless of orientation.
Consequently, these lead to numerical results around $10^{-16}$ or less, for
both methods, and were therefore not represented in
Fig.~\ref{fig:comparison_Francois_me}. Appendix~\ref{sect:cancelling} explains
why these disk-integration factors cancel out.

\section{Overall visibilities}
\label{sect:photo_CoRoT}

In what follows we will first focus on mode visibilities in the CoRoT
photometric before dealing with amplitude ratios between different bands.

\subsection{General results}

\begin{figure}[p]
\includegraphics[width=\textwidth]{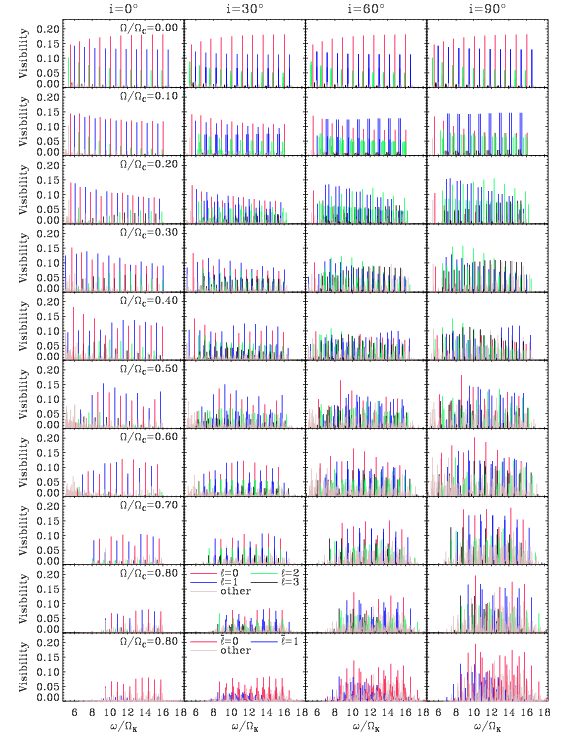}
\caption{(Colour online) Mode visibilities for $2\,M_{\odot}$ stellar models at
$9$ different rotation rates, and $4$ different inclinations ($i=0^{\circ}$
corresponds to a pole-on configuration).  The colours correspond to the $\l$
value carried on from $\Omega = 0$, except for the last row where they
correspond to $\tilde{\l}$.\label{fig:visi}}
\end{figure}

Fig.~\ref{fig:visi} shows the visibilities of modes within acoustic frequency
ranges for a set of $2\,M_{\odot}$ stellar models at $9$ different rotation
rates (1 for each row).  The columns correspond to $4$ different inclinations. 
Given that the intrinsic mode amplitudes cannot currently be predicted,
as this would require a full non-linear development, we normalised the modes so
that the maximal displacement within the entire star,
$\|\vect{\xi}\|_{\mathrm{max}}$, multiplied by the square of the 
co-rotating frequency, $(\omega + m\Omega)^2$, is
constant.  This tends to favour acoustic modes, for which the maximal amplitude
is near the surface.  Multiplying by the square of the frequency yields a near
constant value for the visibilities of modes with similar $\l$ and $m$ in the
acoustic asymptotic regime, for the non-rotating model (upper row). The
different colours indicate the $\l$ value which would carry on from $\Omega=0$
through a mode-following procedure \citep[\eg\ ][]{Lignieres2006}.  In
\citet{Lignieres2008} \citep[see also][]{Reese2008b}, a different set of quantum
numbers $(\tilde{n},\,\tilde{\l},\,m)$ is introduced.  This set of quantum
numbers is better adapted to island modes, the rapidly rotating equivalent of
modes with low $\l-|m|$ values.  The number $\tilde{n}$ corresponds the number
of nodes along the underlying ray path associated with these modes, whereas
$\tilde{\l}$ is the number of nodes transverse to this ray path.  In the last
row of Fig.~\ref{fig:visi}, the visibilities of the $0.8\,\OmegaC$ model are
repeated, but using a colour scheme which indicates the value of $\tilde{\l}$.

A first trend which appears at rapid rotation rates is that the equator-on
configurations tend to have larger amplitudes.  This is simply due to the fact
that we normalise the visibilities with respect to the stellar luminosity in the
observer's direction.  Given that the star is less luminous equator-on, this
leads to larger relative amplitudes.  If instead we normalise by the pole-on
luminosity, such as what is done in \citet{Lignieres2006}, the amplitudes show a
much smaller increase, due to the equatorial focusing of island modes, as will
be discussed in the following section.

A second effect which can be seen in these panels is that the pulsation spectra
become more and more complex as $\Omega$ increases.  This is caused by several
factors.  First of all, as opposed to the non-rotating case, modes with the same
$n$ and $\l$ values but different $m$ values have different frequencies.  At
small (uniform) rotation rates, this leads to frequency multiplets which are
evenly spaced and which span a small frequency range.  As the rotation rate
increases, these multiplets become uneven and they start to overlap, thereby
leading to a complex spectrum.  Secondly, modes with intermediate $\l-|m|$
values become chaotic modes at rapid rotation rates.  This causes their node
spacing to become uneven and is likely to make cancellation effects less
efficient than in the non-rotating case \citep{Lignieres2009}.  Accordingly, the
number of modes which are visible above a given threshold increases.  One may
wonder if this can explain the irregular nature of pulsation spectra in $\delta$
Scuti stars.  Up to some extent yes, but not entirely.  In both cases, the
frequencies are distributed so that no simple regularities stand out.  However,
observations show a somewhat larger dispersion in amplitudes, with only a few
high-amplitude modes.  This suggests that non-linear effects lead to intrinsic
mode amplitudes which cannot be described in a simple way as is done here with
the ad-hoc normalisation, thus making it difficult to identify modes from
visibilities alone.

Another important effect is the relative simplicity of the frequency spectra for
the pole-on orientation, and to a lesser extent for $i=30^{\circ}$, even in the
most rapidly rotating models.  Indeed, non-axisymmetric modes cancel out in the
pole-on configuration, and the $\tilde{\l}=0$ modes, \ie\ the rotating
counterparts to $\l=0$ and $1$ modes, stand out. This means that such an
orientation facilitates interpreting the oscillation spectrum of rapidly
rotating stars.  Of course, it is also more difficult to identify pole-on stars
as rapid rotators, since spectroscopic observations will only reveal narrow
absorption lines due to the unknown $\sin i$ factor (although the overall shape
of the spectrum will differ substantially from that of a black-body at a single
temperature).  Interferometric studies have, however, been able to confirm that
Vega is a rapidly rotating star, seen nearly pole-on \citep[$i=4.54^{\circ}$,][]
{Peterson2006b}.  This recently motivated a search for pulsation modes in this
star.   \citet{Boehm2012} found some stellar variability, but were unable to
confirm the presence of stellar oscillations, thereby showing the need for
further observations with larger instruments.  It is also necessary to search
for other pole-on rapid rotators to see if they exhibit stellar pulsations.

\begin{figure}[htbp]
\includegraphics[width=\textwidth]{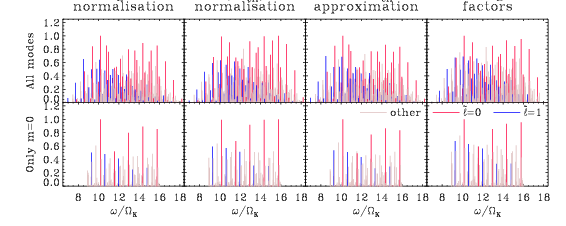}
\caption{(Colour online) Mode visibilities calculated or normalised in alternate
ways for the model rotating at $\Omega = 0.8\OmegaC$, as viewed from the
equator.  The upper shows the visibilities of all of the modes whereas the lower
panel only includes axisymmetric $(m=0)$ modes.  See text for
details.\label{fig:visi_tests}}
\end{figure}

In the equator-on configurations, the visibilities of the chaotic modes
increases relatively to those of the island modes, thereby making it harder to
distinguish the two.  This result partially agrees with the results presented in
\citet{Lignieres2009}, although we do note that here, some of the island modes
still tend to remain more visible whereas in \citet{Lignieres2009} the two had
comparable amplitudes.  To understand where this difference comes from, we tried
various manipulations.  In Fig.~\ref{fig:visi_tests}, we show alternate ways of
calculating or normalising the visibilities for the model rotating at $\Omega =
0.8\,\Omega_{\mathrm{C}}$.  The top row shows the visibilities of all of the
modes, whereas the bottom row corresponds to axisymmetric modes $(m=0)$ only.
The first column uses the original normalisation, so that the upper left panel
of Fig.~\ref{fig:visi_tests} is identical to the lower right panel of
Fig.~\ref{fig:visi}. The second column shows shows what happens if the modes are
normalised by $\left< \delta \Teff \right>$ rather than $\omega^2
\|\vect{\xi}\|_{\mathrm{max}}$.  The third column shows what happens if we use
the approximation $\delta T \simeq \delta \Teff$ instead of $\delta T/T \simeq
\delta \Teff/\Teff$.  Finally, the last column shows the disk-integration
factors as given in Eq.~(\ref{eq:visibility_francois}).  For the sake of
legibility, we divided the visibilities in each panel by the maximal value, so
that the results are between $0$ and $1$.  As can be seen, the most visible
modes still tend to be island modes regardless of how the visibilities are
calculated and whether or not only axisymmetric modes are kept.  Hence, another
explanation is needed for the difference between the present results and those
of \citet{Lignieres2009}.  One difference between the two studies is the
boundary condition on the pressure perturbation.  Indeed, in the present study,
we set the vertical gradient of $\delta p/P_0$ to zero at the surface
(Eq.~\ref{eq:mechanical_bc}), in order to avoid having $\delta T = 0$ at the
surface.  \citet{Lignieres2009} set $\delta p = 0$ instead. This
however does not lead to $\delta T = 0$ in their case, since they are dealing
with a polytropic model, where the pulsation equations become singular at the
surface.  This difference in boundary condition may have an important impact on
the temperature fluctuations at the surface which dominate the mode visibilities
presented here, although it is hard to test since applying the condition
$\delta p = 0$ with SCF models would only lead to $\delta T = 0$.  One
possibility would be to apply the condition
Eq.~(\ref{eq:mechanical_bc}) to polytropic models.

\begin{figure}[htbp]
\includegraphics[width=\textwidth]{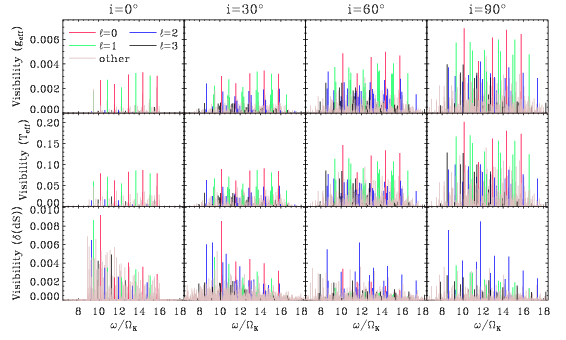}
\caption{(Colour online) Separate contributions to the visibilities, coming from
the $\geff$, $\Teff$ and the $\vect{\mathrm{d}S}$ variations.  These are calculated for
the model at $\Omega=0.8\,\OmegaC$ (which corresponds to the before last row of
Fig.~\ref{fig:visi}).\label{fig:visi_separate}}
\end{figure}

Finally, Fig.~\ref{fig:visi_separate} shows the separate contributions coming
from the $\geff$, $\Teff$ and $\vect{\mathrm{d}S}$ variations.  As can be seen in the figure,
the effective temperature fluctuations are the most dominant effect.  This is
because effective gravity plays a small role in the specific intensity emitted
by an atmosphere.  The surface deformation plays a small role in acoustic modes
because it is proportional to the displacement rather than the displacement
times the square of the frequency.

\subsection{Equatorial concentration}

One of the consequences of the centrifugal force on acoustic modes is the
equatorial focusing of island modes around the equator, due to the geometric
distortion of the resonant cavity.  \citet{Lignieres2006} pointed out the
possibility that this could lead to higher observed pulsation amplitudes in
stars observed in an equator-on configuration, thereby explaining the
observations of \citet{Suarez2002}, which show a correlation between mode
amplitude and inclination.  It is also worth noting that for gravito-inertial
modes in the inertial regime, modes become trapped in an equatorial waveguide
\citep[\eg][]{Dintrans2000,Townsend2003a}.  Paradoxically, this usually causes
these modes to be more visible in a pole-on configuration, due to the
contribution from surface distortion \citep{Townsend2003b,
Daszynska_Daszkiewicz2007}.

\begin{figure}[htbp]
\includegraphics[width=\textwidth]{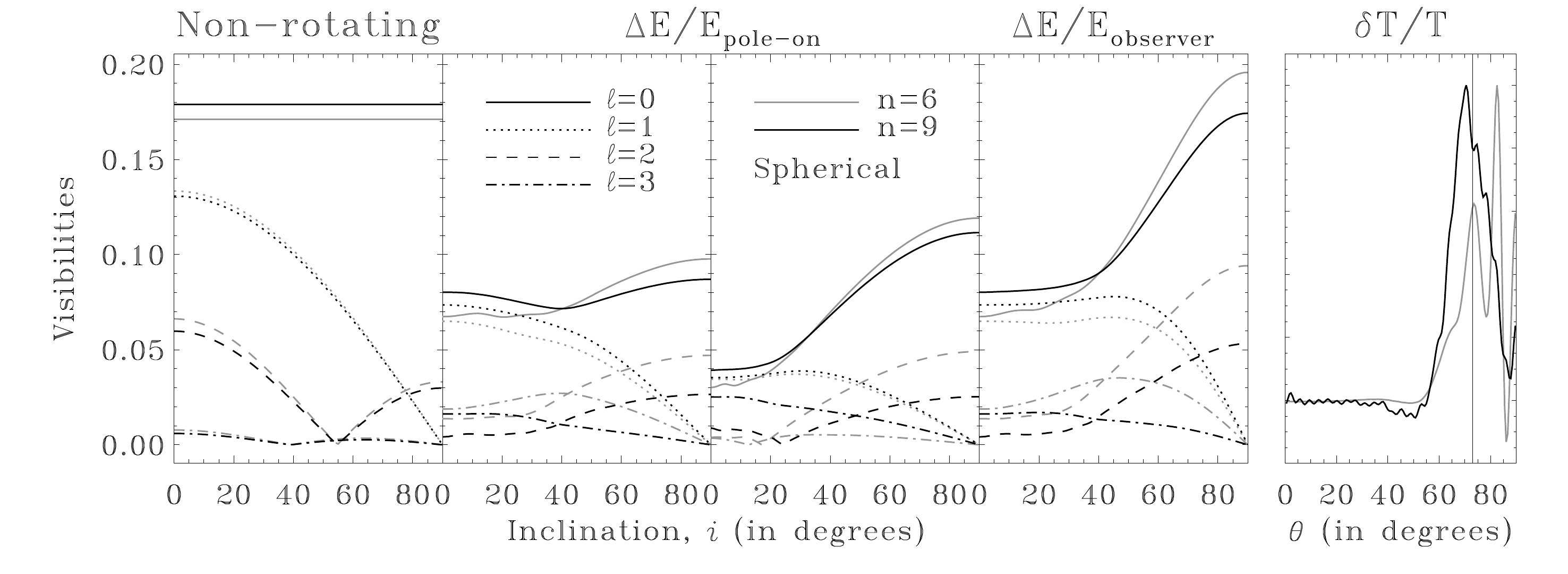}
\caption{Visibilities of various island modes as a function of the stellar
inclination.  The $\l$ and $n$ values are indicated by the styles and shades of
the lines.  The rightmost detached panel shows the temperature fluctuations,
$\delta T/T$, of the two $\l=0$ modes in the rotating
model.\label{fig:equatorial_concentration}}
\end{figure}

Figure~\ref{fig:equatorial_concentration} shows the visibilities of various
axisymmetric ($m=0$) acoustic modes as a function of inclination. The left panel
corresponds to a non-rotating model, whereas the three middle panels corresponds
to a model at $0.8\,\OmegaC$.  At such a rotation rate, all of the modes in
these panels are island modes.  The mode visibilities in the second panel are
normalised by the pole-on equilibrium luminosity of the star \citep[as is done
in][]{Lignieres2006}.  What is surprising in this figure is that although the
symmetric modes are more visible from the equator than from the poles, one
would expect a higher contrast between the two, especially for the $\l=0$
modes.  Indeed, the surface temperature fluctuations, $\delta T/T$, of the
$\l=0$ modes, depicted in the rightmost detached panel, show a much stronger
confinement towards the equatorial region.  A careful investigation reveals that
the reason for this lack of contrast between polar and equatorial visibilities
stems from the geometrical shape of the star.  Given the high curvature at the
equator, the surface quickly starts pointing more towards a polar direction.  To
illustrate this point more clearly, the third panel shows the same visibilities
assuming the star is spherical, but retaining the other effects of rotation such
as gravity darkening.  Clearly, a much stronger contrast appears between the
poles and the equator.  One can also quantify at what latitude a temperature
fluctuation contributes more to mode visibilities in an equatorial direction
than in a polar one by calculating at what point the ratio,
\begin{equation}
R\left(\theta_0\right) = \frac{2\int_{\phi=-\pi/2}^{\pi/2} \dpart{I}{\Teff}\left[\vect{n}\left(\theta_0,\phi\right)\cdot \vect{e}_x,
                               \geff\left(\theta_0\right),\Teff\left(\theta_0\right)\right] \vect{e}_x \cdot \vect{\mathrm{d}S}}
                              {\int_{\phi=-\pi}^{\pi} \dpart{I}{\Teff}\left[\vect{n}\left(\theta_0,\phi\right)\cdot \vect{e}_z,
                               \geff\left(\theta_0\right),\Teff\left(\theta_0\right)\right] \vect{e}_z \cdot \vect{\mathrm{d}S}},
\end{equation}
exceeds one.  In the above formula, only the effects of temperature fluctuations
are retained as they usually are the dominant factor in mode visibilities.  The
denominator corresponds to the intensity fluctuation in the polar direction,
$\vect{e}_z$, caused by a unit axisymmetric effective temperature variation at
colatitude $\theta_0$.  The numerator is the intensity fluctuation in the
equatorial direction, $\vect{e}_x$, caused by the same unit effective
temperature fluctuation, and it's mirror image in the southern hemisphere, at
colatitude $\pi-\theta_0$, hence the factor $2$. In the rotating model used in
Fig.~\ref{fig:equatorial_concentration}, the ratio $R(\theta_0)$ exceeds $1$
from $73.1^{\circ}$, which is depicted by the vertical line in the rightmost
panel.  As can be seen in the figure, this line is quite close to the maximum of
the $(n,\,\l)=(9,0)$ mode, thereby explaining the near equal pole-on and
equator-on visibilities. Had the star been spherical, the position of this line
would have been $57.0^{\circ}$.

The fourth panel shows the mode visibilities if they are normalised by the
equilibrium luminosity in the observer's direction (as is done in
Fig.~\ref{fig:visi}).  Once more, there is a strong contrast between the polar
and equatorial visibilities.  This is simply caused by the lower luminosity of
the star equator-on, which results from gravity darkening.  From an
observational point of view, this last ratio is easier to obtain given that
$E_{\mathrm{pole-on}}$ is not directly observable, and may be one of the
dominant factors in the correlation between inclination and mode amplitude found
by \citet{Suarez2002}.

\subsection{Avoided crossings}
\label{sect:avoided_crossings}

It is then interesting to look at the effects of avoided crossings on mode
visibilities. \citet{Lignieres2006} pointed out the possibility that
rotationally-induced avoided crossings may explain close frequency pairs
observed in $\delta$ Scuti stars \citep{Breger2002, Breger2006b}.  Indeed, the
structure of the eigenfunctions of the two (or more) modes involved in an
avoided crossing are a mixture of the eigenfunctions prior to the crossing. 
Hence, if a high-visibility mode interacts with a low-visibility mode, the two
may be visible during the crossing when both modes have a mixed character, as
was already shown in \citet{Daszynska_Daszkiewicz2002}.

\begin{figure}[htbp]
\includegraphics[width=0.38\textwidth]{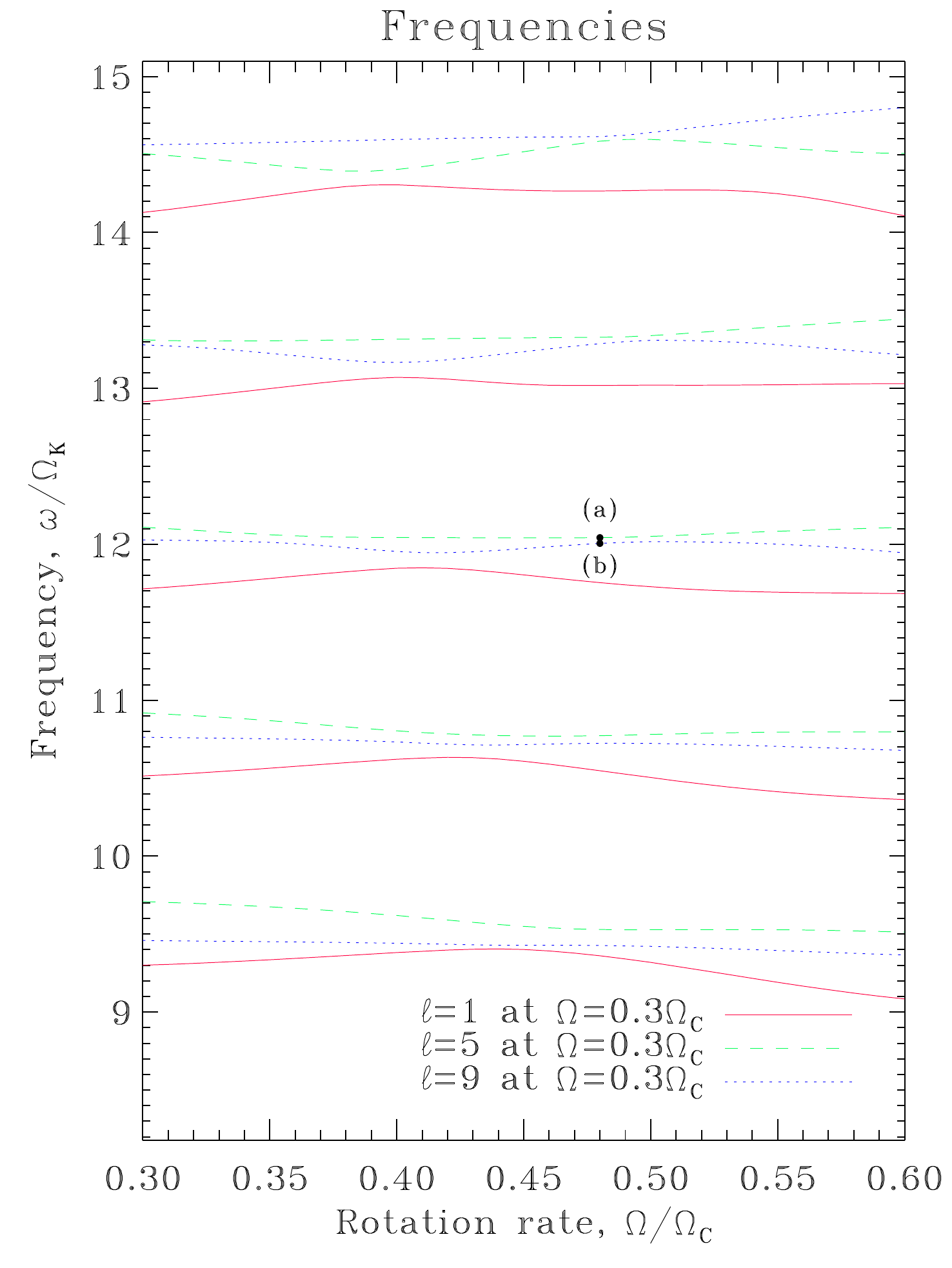}
\includegraphics[width=0.38\textwidth]{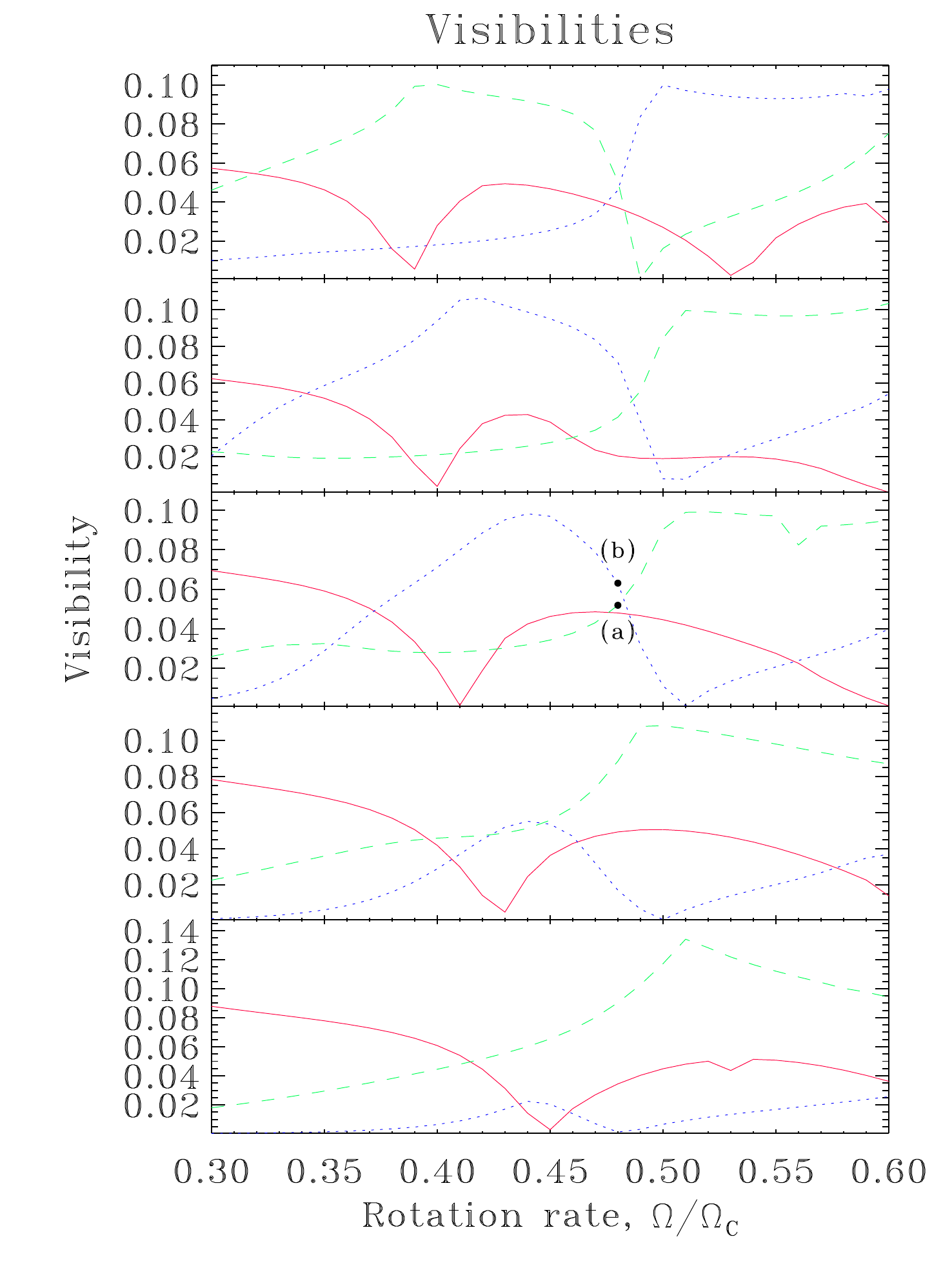} 
\includegraphics[width=0.20\textwidth]{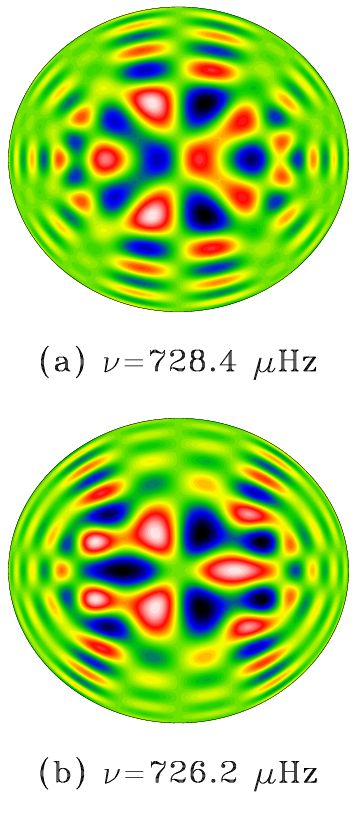}
\caption{(Colour online) A set of avoided crossings between $\l=1$, $5$ and $9$
modes with $m=1$.  The linestyles and colours indicate the $\l$ value at
$\Omega=0.3\,\OmegaC$ (note that an avoided crossing has already taken place
between the uppermost $\l=5$ and $9$ modes, hence the ``swapped'' identification
compared to the other modes).  The left panel shows the frequencies, the middle
panel gives the visibilities in the CoRoT photometric band for $i=30^{\circ}$
and the right panel shows the meridional cross-sections of two modes undergoing
an avoided crossing, and labelled by ``(a)'' and ``(b)'' in the two left panels.
\label{fig:avoided_crossings}}
\end{figure}

Figure~\ref{fig:avoided_crossings} (left panel) shows a set of avoided crossings
between families of $\l=1$, $5$ and $9$ modes, the azimuthal order being $m=1$. 
As can be seen in the figure, rotation is an ideal way of generating avoided
crossings due to its differential effect on modes of different degrees. As was
found in \citet{Lignieres2006}, we also find that modes of degree $\l$ tend to
interact with modes of degree $\l+4$.  The middle panel shows the associated
mode visibilities.  At $0.48\,\OmegaC$, the modes labelled (a) and (b) are
strongly interacting, thereby producing an intermediate visibility for both
modes.  We note that the frequency separation between the two is
$2.24\,\mu\mathrm{Hz}$, which is slightly larger than for observed close
frequency pairs.  The meridional cross-section of these modes is depicted in the
right panel, both of which show mixed characteristics.

In order for avoided crossings to explain close frequency pairs, a number of
conditions need to be meet.  On the one hand, the coupling between the modes
needs to be sufficiently weak so that the frequency differences become small
enough.  In the above example, none of the frequencies get closer than
$1.4\,\mu\mathrm{Hz}$ which is slightly above what is considered to be a close
frequency pair.  On the other hand, the coupling needs to be sufficiently strong
so as to allow the modes to have mixed characteristics over a non-negligible
range in parameter space.  Then, one would need to do a full statistical study of
the frequency spectrum of visible modes and quantify how many close frequencies
are due to avoided crossings and how many result from chance coincidence between
uncoupled modes.  Such a study is, however, beyond the scope of this paper.

\section{Multi-colour visibilities}
\label{sect:photo_multi}

One of the advantages of calculating amplitude ratios between different
photometric bands is that the intrinsic amplitude of the mode factors out.  As a
result, it is possible to predict such ratios from linear theory alone.  In this
section, we will examine some of the effects of rotation on amplitude ratios. 
In particular, we will focus on a frequency multiplet, and on modes with the
same $\l$ and $m$ values.  However, before discussing different results, it is
useful to examine more carefully the question of normalising mode visibilities
in multiple photometric bands.

\subsection{Normalisation}

In what follows, we will depart from the traditional approach which consists in
normalising the visibilities with respect to a given band. The problem with such
an approach is that sometimes, there are modes for which their visibilities
nearly cancel out in the chosen band, thereby amplifying their amplitude
ratios.  This can then exaggerate differences between these ratios and those of
other modes.  We therefore propose another normalisation scheme which does not
favour one band over the others, and which applies to a set of modes which we
want to compare.

We start by considering a set of $N$ modes, numbered $j=1 \dots N$.  Let us
denote $V_i^j$ the visibility of mode $j$ in the photometric band $i$.  We will
also use the notation $\vect{V}^j$ to represent the vector $(V_i^j)_{i=1 \dots
b}$.  We want to normalise each mode so as to minimise, in some sense, the
distance between their visibilities.  To do so, we introduce a set of references
visibilities $W_i$, which at this point are unknown, and normalise each mode so
as to minimise its distance to $W_i$.  This amounts to minimising the following
cost function:
\begin{equation}
J = \sum_{j=1}^{N} \sum_{i=1}^{b} \left( A_j V_i^j - W_i \right)^2,
\label{eq:cost_simple}
\end{equation}
where $b$ is the number of photometric bands and $A_j$ the normalisation factors for
each mode.  In this expression, both the $A_j$ and $W_i$ are unknowns. 
Furthermore, we impose the following constraint, so as to avoid the trivial
solution $A_j \equiv 0$, $W_i \equiv 0$:
\begin{equation}
\sum_{i=1}^{b} W_i^2 = 1.
\end{equation}
As will be shown in App.~\ref{app:normalisation}, $\vect{W} = (W_i)_{i=1
\dots b}$ turns out to be the principal component of the normalised vectors
$\left(\tilde{\vect{V}}^j \right)_{j=1 \dots N}$, where $\tilde{V}^j_i =
V^j_i/\sqrt{\sum_{k=1}^{b} \left(V^j_k\right)^2}$.  This result is fairly
intuitive, since the vectors $\left(\tilde{\vect{V}}^j \right)_{j=1 \dots N}$ retain
the directional information (in $b$ dimensions) associated with each mode while
discarding the amplitude (which is arbitrary anyway).  Once the reference
visibilities are determined, it is straightforward to find the normalisation
factors:
\begin{equation}
A_j = \frac{\sum_{i=1}^{b} V_i^j W_i}{\sum_{i=1}^{b} \left(V_i^j\right)^2},
\qquad \mbox{or equivalently,} \qquad \tilde{A}_j = \sum_{i=1}^{b} \tilde{V}_i^j W_i.
\label{eq:Aj}
\end{equation}
The normalised visibilities then become $\hat{\vect{V}}^j = A_j \vect{V}^j =
\tilde{A}_j \tilde{\vect{V}}^j$. This last equation has a simple geometrical
interpretation: indeed, $\tilde{A}_j \tilde{\vect{V}}^j$ represents the
projection of $\vect{W}$ onto the direction
$\vect{\tilde{V}}^j$.

In the following sections, we plot normalised visibilities $\hat{\vect{V}}^j$
based on the set of modes relevant to each sub-panel of the various figures, and
will call them ``amplitude ratios'' even if strictly speaking they are not
amplitude ratios. In some cases, the reference visibilities, $W_i$, are also
displayed.  It is important to bear in mind that this normalisation depends on
the set of modes, and furthermore on their inclination since this affects their
visibilities in rotating stars.  Hence, the mode-dependant normalisation factors
will change from one panel to another, even for the same set of modes, if viewed
at different inclinations. Arguably, some of the information based on
inclination is lost in the process (the same would also be true if we had
normalised the visibilities with respect to a given band).  However, what is
important here is the comparison of amplitude ratios between a set of modes for
a given configuration.  Furthermore, one does not have the luxury to modify the
inclination of an observed star. Another concern is that by normalising the
visibilities by a mode-depend scale factor, one is seemingly no longer comparing
amplitude ratios which are the true invariants.  However, such information is
contained within the shape of the curve defined by the normalised visibilities,
\ie\ the ratios between the components $(\hat{V}_i^j)_{i=1\dots b}$.  Moreover,
with the above normalisation, modes with similar amplitude ratios will have
similar scale factors, $\tilde{A}_j$, and hence normalised visibilities which
are similar, even more so than if one of the photometric bands had been used for
the purposes of normalisation.

\subsection{A multiplet}

\begin{figure}[htbp]
\includegraphics[width=\textwidth]{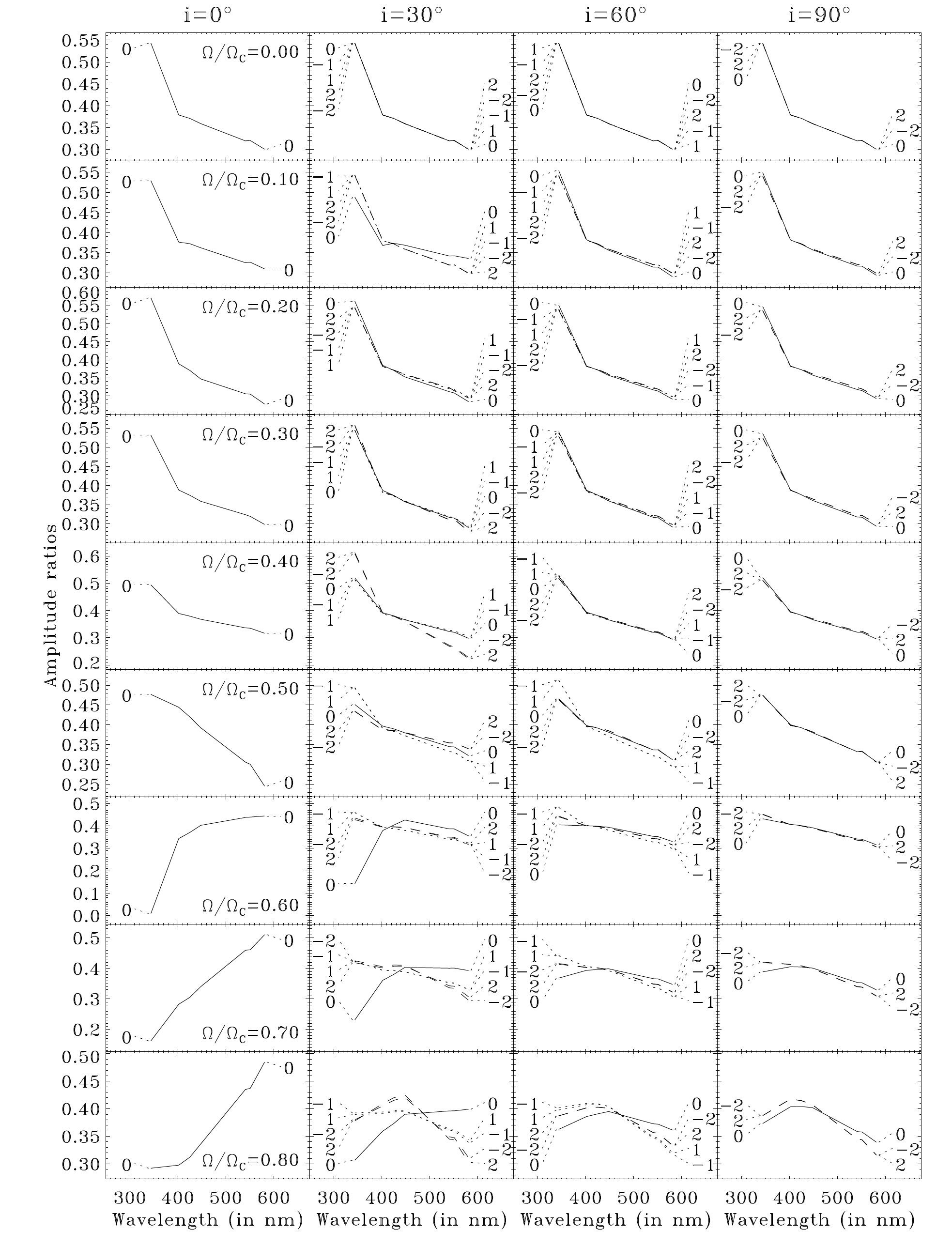}
\caption{Amplitude ratios of the $(n,\,\l) = (6,2)$ multiplet, based on the 
Geneva photometric system.  Each column corresponds to a different orientation
and each row to a different rotation rate.  The numbers on either end of each
plot and connected by dotted lines indicate the azimuthal order, $m$.  The
continuous, dotted, and dashed  linestyles correspond to $m=0$, $|m|=1$ and
$|m|=2$, respectively. \label{fig:ratios}}
\end{figure}

We first start by looking at the $(n,\,\l)=(6,2)$ multiplet.
Figure~\ref{fig:ratios} shows amplitude ratios, based on the Geneva photometric
system, at different rotation rates and inclinations. Only axisymmetric modes
are displayed for the pole-on configurations $(i=0^{\circ})$ and even modes,
\ie\  modes which are symmetric with respect to the equator, are shown for the
equator-on configurations  $(i=90^{\circ})$, given that the other modes cancel
out.  As expected, the amplitude ratios do not depend on the inclination or
azimuthal order in the non-rotating case.  When the rotation rate increases,
this is no longer true, as has been already pointed out by
\citet{Daszynska_Daszkiewicz2002} and \citet{Townsend2003b}.  It is interesting
to note that this departure from a constant behaviour is non-monotonic and
therefore, somewhat difficult to predict. For instance, at
$(\Omega,i)=(0.4\,\OmegaC,30^{\circ})$ the $|m|=2$ modes have higher ratios at
lower wavelengths and lower ratios at higher wavelengths.  At
$(\Omega,i)=(0.5\,\OmegaC,30^{\circ})$ the opposite is true.  The $|m|=1$ modes
have the opposite behaviour.  The dependence on orientation seems to be somewhat
simpler, although one can note how the mode order frequently changes between
$i=30^{\circ}$ and $i=60^{\circ}$.  This implies that mode identification from
amplitude ratios will not be as straightforward as in the non-rotating case.

\begin{figure}[htbp]
\includegraphics[width=\textwidth]{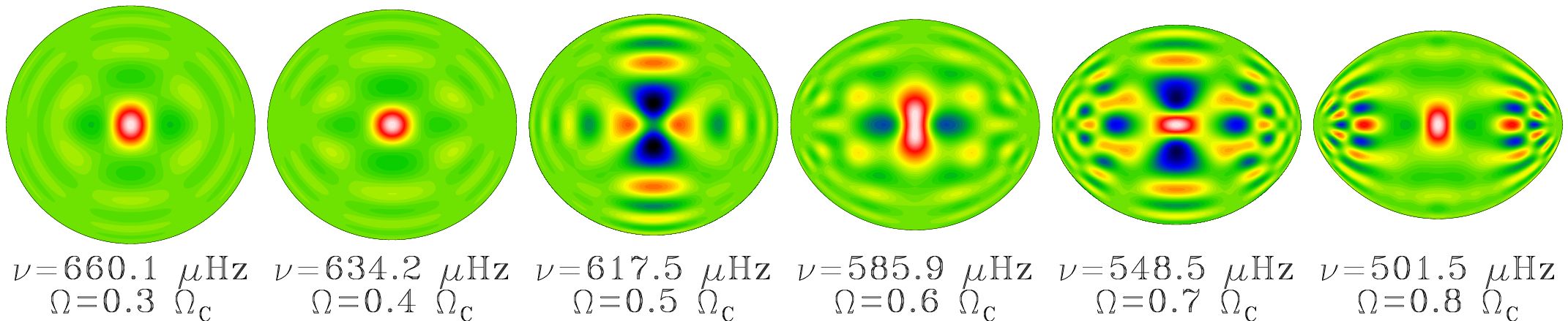}
\caption{Meridional cross sections of the $(n,\,\l,\,m)=(6,2,0)$ mode at
different rotation rates.  At the lowest rotation rates, the modes' structure is
dominated by the central parts. \label{fig:fields_ratios}}
\end{figure}

It is also interesting to note how the overall shape of most of the ratios
change starting from $\Omega = 0.6 \,\OmegaC$.  At lower rotation rates, the
concavity seems to be pointing up and the largest ratios are obtained at the
lowest wavelength.  In the most rapidly rotating models, the concavity is
pointing downwards, and the ratios at the lowest wavelength are among the
smallest.  A particularly striking example of a change in behaviour is the
drastic modification of the $m=0$ amplitude ratios between $\Omega = 0.5
\,\OmegaC$ and $\Omega = 0.6 \,\OmegaC$, especially for the pole-on
configurations $(i=0^{\circ})$.  In order to gain a better understanding of what
is causing this, we plot in Fig.~\ref{fig:fields_ratios} the meridional
cross-sections of this mode for a selection of rotation rates. As can be seen,
an avoided crossing takes place between $\Omega = 0.5 \,\OmegaC$ and $\Omega =
0.6 \,\OmegaC$.  This agrees with \citet{Daszynska_Daszkiewicz2002} who had also
shown that avoided crossings can have an important effect on amplitude ratios
and phase differences, and who furthermore showed how these can lead to
erroneous identifications.  A number of other avoided crossings involving the
non-axisymmetric modes also take place around that rotation rate.  Another effect
which comes into play is the fact that the modes depart from the polar regions
to focus around the equator starting from $\Omega \simeq 0.6\,\OmegaC$, as can
be seen in Fig.~\ref{fig:fields_ratios} for the $m=0$ modes.

One last interesting feature in Fig.~\ref{fig:ratios} is the similarity
between prograde and retrograde modes with the same $|m|$ value. 
\citet{Townsend2003b} had previously noted a similar phenomena for
gravito-inertial modes, using the traditional approximation.  He was, however,
unable to provide an explanation for this behaviour given that the Coriolis
force plays a dominant role in the modes he studied, thereby leading to
important differences between prograde modes and their retrograde counterparts. 
The modes shown here are acoustic modes, in which the Coriolis force only plays
a minor role.  Hence, the structures of prograde modes and their retrograde
counterparts are quite similar, thus explaining the similar amplitude ratios.

\subsection{Modes with the same $\l$ and $m$ values}

\begin{figure}[htbp]
\includegraphics[width=\textwidth]{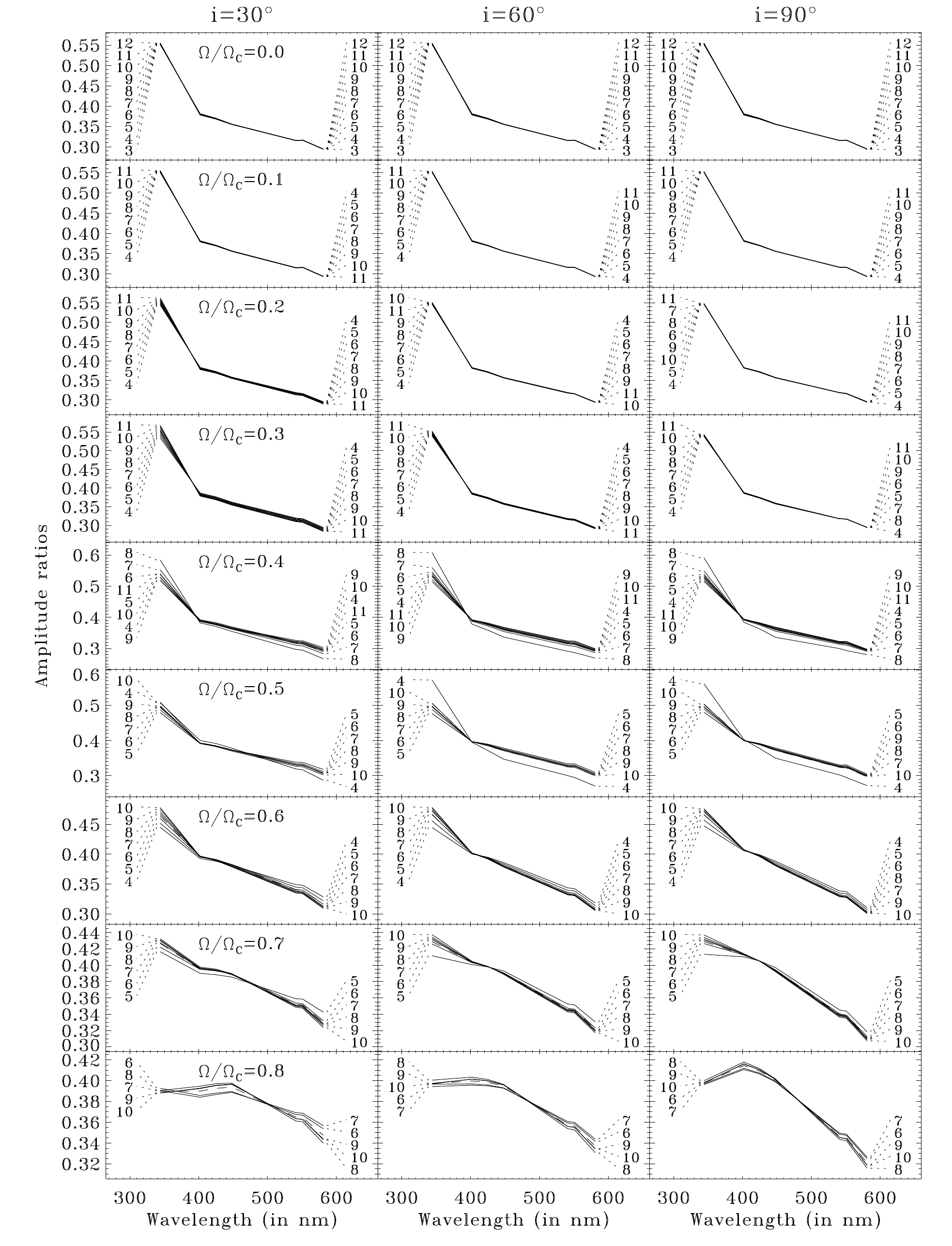}
\caption{Amplitude ratios for sets of modes with consecutive radial orders and
for which $(\l,\,m) = (1,1)$.  The reference visibilities used for normalisation,
$W_i$, are shown as a dashed line in each plot.  In most cases, this line is
covered up by the amplitude ratios, which are represented by continuous lines. 
The numbers on either end of the plots indicate the radial orders, $n$.
\label{fig:ratios_l_m}}
\end{figure}

Figure~\ref{fig:ratios_l_m} shows the amplitude ratios for a set of modes with
consecutive radial orders, $n$, and with $(\l,\,m) = (1,1)$. In the non-rotating
case, amplitude ratios for fixed $(\l,\,m)$ values are expected to be similar,
though not identical, since the ratios between $\hat{\xi}_r/r$, $\delta \hTeff/\Teff$,
and $\delta \hgeff/\geff$ can vary slightly from one mode to the next.  This is
different from the purely geometrically disk-integration factors calculated in
\citet{Lignieres2006} where strict equality is expected (and obtained) at
$\Omega = 0$.  What is interesting to note here is that for the most part, this
similarity between the different ratios continues up to rapid rotation rates, as
can be seen in Fig.~\ref{fig:ratios_l_m}.  The reason for this behaviour is
straightforward.  Modes with fixed $(\l,\,m)$ values produce island modes with
the same $(\tilde{\l},\,m)$ values.  These modes have a similar lateral
structure perpendicular to the underlying ray path, apart from the width which
decreases with frequency, and only differ by the number of pseudo-radial nodes. 
Recently, \citet{Pasek2012} came up with asymptotic expressions for their
structure, based on ray-dynamics.  One may then wonder why we didn't directly
select modes according to their $(\tilde{\l},\,m)$ values.  The problem with
selecting modes that way is that it mixes together even and odd modes, which
have different visibilities when the star is close to equator-on.  Although the
lateral structure is the same, the underlying ray path curves around the equator
so that in odd modes, one of the pseudo-radial nodes actually corresponds to the
equator.  Hence, even and odd modes with the same $(\tilde{\l},\,m)$ values will
have a similar surface structure, \textit{if one limits themselves to one of the
hemispheres}.  Accordingly, one set of $(\tilde{\l},\,m)$ values correspond to
\textit{two} different sets of $(\l,\,m)$ values, one for even modes and the
other for odd modes.

\begin{figure}[htbp]
\includegraphics[width=\textwidth]{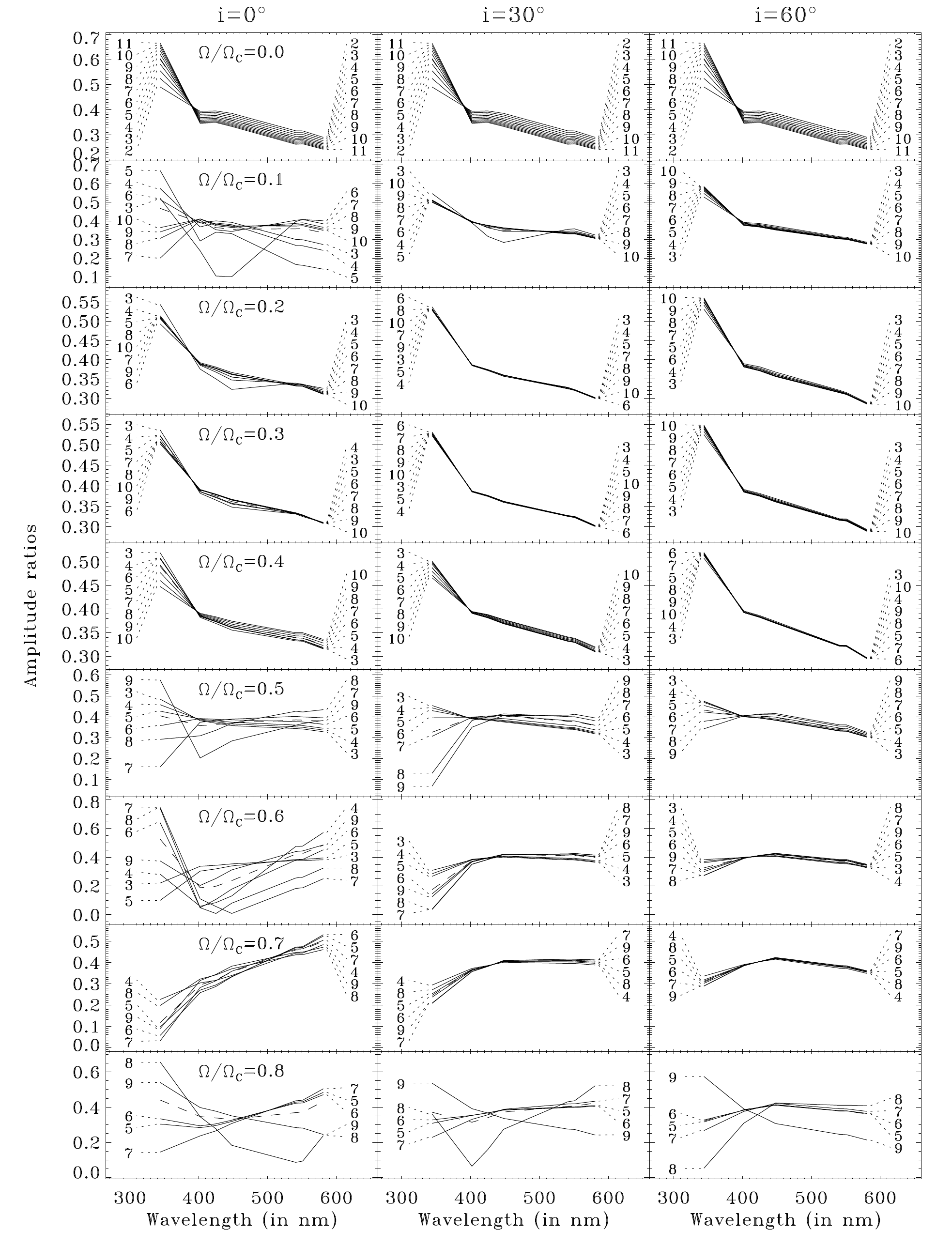}
\caption{Same as Fig.~\ref{fig:ratios_l_m}, except that
$(\l,\,m)=(3,0)$.\label{fig:ratios_l_m_bis}}
\end{figure}

It would, however, be incorrect to conclude that modes with the same $\l$ and
$m$ values systematically lead to similar amplitude ratios.  In
Fig.~\ref{fig:ratios_l_m_bis}, we represent the amplitude ratios for modes with
$(\l,\,m)=(3,0)$.  As opposed to Fig.~\ref{fig:ratios_l_m}, this is one of the
worst cases, where agreement is rather poor.  Various effects can lead to
differences in amplitude ratios.  As mentioned above, avoided crossings can
cause irregular behaviour.  In some cases, a mode can be at the transition where
temperature variations start to dominate over surface deformations.  This can
lead to a very different behaviour from the modes before or after the
transition, and cause some of the visibilities to nearly cancel out.  A good
example of this is the $n=6$ mode for $i=0^{\circ}$ and $\Omega=0.1\OmegaC$. 
Finally, if the radial order is too low, a given mode may depart sufficiently
from the asymptotic regime to produce noticeable differences on the
visibilities.

\section{Conclusion}

In this paper, we derived a new set of equations to calculate mode visibilities
in rapidly rotating stars.  These equations take into account the centrifugal
deformation of the star as well as gravity darkening, and treats the modes in a
more realistic way by including the associated effective temperature and gravity
variations, as well as the surface distortions.  The modes are calculated in
fully deformed 2D stellar models based on the SCF method
\citep{Jackson2005,MacGregor2007}, using the 2D oscillation code TOP
\citep{Reese2006, Reese2009a}.  A grid of Kurucz atmospheres was used to
calculate realistic intensities at each point on the surface
\citep{Kurucz2005}.  As such, this represents an important step towards
obtaining realistic multi-colour visibilities of acoustic modes in rapidly
rotating stars, and an improvement over previous studies which approximated the
effects of rotation on the model and the oscillations
\citep[\eg][]{Daszynska_Daszkiewicz2002,Townsend2003b} or used simplified
visibility calculations \citep{Lignieres2009}.

One of the important limitations in the present study is the use of the
adiabatic approximation.  Accordingly, the temperature variations are not
reliable near the surface and, hence, provide a poor approximation of the
effective temperature variations \citep{Dupret2002}.  Including non-adiabatic
effects is likely to have an important impact on mode visibilities since
effective temperature variations often play a dominant role in mode
visibilities, as was shown in Fig.~\ref{fig:visi_separate}.  Nonetheless, one
can hope that the present calculations will give insight into the qualitative
behaviour of mode visibilities and amplitude ratios in rapidly rotating stars. 
In future studies, we plan to implement non-adiabatic effects, including a full
treatment of the stellar atmosphere, by applying a similar approach to
\citet{Dupret2002}.  This will require the use of thermally relaxed rapidly
rotating models, such as those produced by the ESTER code \citep{Rieutord2009,
Espinosa2010}.

The results presented here confirm a number of results previously established.
For instance, in stars observed pole-on, some of the island modes stand out
compared to other modes and form a regular frequency pattern
\citep{Lignieres2009}. Consequently, rapidly rotating pulsating stars seen
pole-on may be promising asteroseismic targets.  Chaotic modes are more
visible than their non-rotating counterparts, probably as a result of irregular
node placement, thereby complicating the frequency spectrum, especially for
equator-on configurations \citep{Lignieres2009}.  Avoided crossings can
substantially modify mode visibilities and cause modes which are not usually
visible to exceed the detection threshold \citep{Daszynska_Daszkiewicz2002}.
\citet{Lignieres2006} previously pointed out that these may explain observed
close frequency pairs in $\delta$ Scuti stars, although a full statistical study
is required to test this possibility.  Finally, amplitude ratios depend on both
the inclination, $i$, and the azimuthal order, $m$, in rotating stars
\citep{Townsend2003b, Daszynska_Daszkiewicz2007}.  This, of course, makes mode
identification in such stars more difficult.

New results were also obtained.  In particular, although rapid rotation causes
island modes to focus around the equator, the corresponding visibilities show a
much smaller contrast between pole-on and equator-on configurations (unless
normalised by star's luminosity in the observer's direction).  This is because
the same geometrical distortion which causes mode focusing in the first place,
also increases the proportion of the stellar surface which points in a more
poleward direction.  We also showed that acoustic modes with the same
$(\l,\,|m|)$ values tend to have similar amplitude ratios, although this effect
is not systematic.  The similarity between prograde and retrograde modes stems
from the small influence of the Coriolis force on acoustic modes.  Modes with
the same $(\l,\,m)$ values have a similar surface structure, as expected from
asymptotic ray theory \citep{Pasek2012}.

In a forthcoming study, we plan to study the statistical properties of the
frequency spectra by analysing their autocorrelation functions
\citep{Lignieres2010} and by looking at the cumulative distribution functions of
the frequency separations.  These will be compared with observations, in order
to assess up to what extent the present theory is realistic and to see if it is
possible to extract global quantities such as the large frequency separation or
rotation rate from the observations.  Using multi-colour photometry, we will also
develop a new strategy for constraining mode identification and obtaining or
confirming the values of global quantities.

\begin{acknowledgements}
We thank the referee, whose comments helped us clarify the manuscript.
We thank M.-A. Dupret, E. Michel, R. Townsend, F. Lignières, and F. Paletou for
interesting discussions which have helped to improve this article.  We wish to
thank S. Jackson, A. Skumanich and T.~S. Metcalfe for their contributions to the
Self-Consistent Field method, used for generating rotating stellar models. We
are very grateful to R. Kurucz for making his ATLAS9 code and opacity data
available and open to the scientific community. We wish to thank the KITP at
UCSB for their warm hospitality during the research program ``Asteroseismology
in the Space Age''. DRR is financially supported through a postdoctoral
fellowship from the ``Subside fédéral pour la recherche 2011'', University of
Liège, and was previously supported by the CNES (``Centre National d'Etudes
Spatiales''), both of which are gratefully acknowledged. This work was granted
access to the HPC resources of IDRIS under the allocation 2011-99992  made by
GENCI (``Grand Equipement National de Calcul Intensif'').  The National Center
for Atmospheric Research is a federally funded research and development center
sponsored by the US National Science Foundation.
\end{acknowledgements}
\bibliographystyle{aa}
\bibliography{biblio}
\appendix

\section{Equations in spheroidal coordinates}
\label{sect:spheroidal_equations}

In this section, we derive explicit expressions for the pulsation equations and
the mechanical boundary condition, based on the coordinate system
described in Sect.~\ref{sect:spheroidal_geometry}. However, before giving these
expressions, it is useful to recall a few definitions.  The natural covariant
basis, denoted $(\Ez,\,\Et,\,\Ep)$, is defined via the relation $\vect{E}_i =
\partial_i \vect{r}$, where $i$ stands for $\zeta$, $\theta$ or $\phi$, and
$\vect{r} = r\er$:
\begin{equation}
\label{eq:covariant_basis}
\Ez = \rz \er, \qquad
\Et = \rt \er + r \et, \qquad
\Ep = r\sint \ep.
\end{equation}
Here $(\er,\,\et,\,\ep)$ is the usual spherical basis associated with the
spherical coordinates $(r\,\,\theta,\,\phi)$. The associated dual
(contravariant) basis is defined such that $\vect{E}^i \cdot \vect{E}_j =
\delta^i_j$:
\begin{equation}
\label{eq:contravariant_basis}
\vect{E}^\zeta  = \frac{\er}{\rz} - \frac{\rt\et}{r\rz}, \qquad 
\vect{E}^\theta = \frac{\et}{r}, \qquad 
\vect{E}^\phi   = \frac{\ep}{r \sint}.
\end{equation}
The vector $\vect{E}^{\zeta}$ is perpendicular to surfaces of constant $\zeta$
value, including the stellar surface. As in \citet{Reese2006}, we derive an
alternate basis from $(\Ez,\,\Et,\,\Ep)$ as follows:
\begin{equation}
\label{eq:alternate_basis}
\az = \frac{\zeta^2}{r^2 \rz} \Ez,\qquad
\at = \frac{\zeta}{r^2 \rz} \Et,\qquad
\ap = \frac{\zeta}{r^2 \rz \sint} \Ep.
\end{equation}
In the spherical limit, the alternate basis converges to the spherical basis. 
The Lagrangian displacement is decomposed over the alternate basis as follows:
\begin{equation}
\vect{\xi} = \xicz \az + \xict \at + \xicp \ap.
\label{eq:alternate_components}
\end{equation}
These components are related to the spherical components (see
Eq.~(\ref{eq:spherical_components})) as follows:
\begin{equation}
\xir = \frac{\zeta^2}{r^2} \xicz + \frac{\zeta\rt}{r^2\rz} \xict, \qquad
\xit = \frac{\zeta}{r\rz} \xict, \qquad
\xip = \frac{\zeta}{r\rz} \xicp,
\end{equation}
where superscripts are used with the alternate components, and subscripts with
the spherical components.  Based on the alternate components, the dot product
$\vect{\xi}\cdot\vgeff$ becomes:
\begin{equation}
\vect{\xi} \cdot \vgeff = \frac{\zeta^2}{r^2\rz} \frac{\dz P_0}{\rho_0} \xicz
                        + \frac{\zeta}{r^2\rz} \frac{\dt P_0}{\rho_0} \xict
                        = \Gz \xicz + \Gt \xict,
\end{equation}
where we have introduced the following quantities:
\begin{equation}
\Gz = \frac{\zeta^2}{r^2\rz} \frac{\dz P_0}{\rho_0},\qquad
\Gt = \frac{\zeta}{r^2\rz} \frac{\dt P_0}{\rho_0}.
\end{equation}

\subsection{Pulsation equations}
\label{sect:spheroidal_pulsation_equations}

We now give explicit expressions for the pulsation equations in spheroidal
coordinates.  The continuity equation is:
\begin{equation}
\label{eq:spheroidal_continuity}
0 = \frac{\delta \rho}{\rho_0} + \frac{\zeta^2}{r^2 \rz}\left[\frac{\dz \left(
              \zeta^2 \xicz \right)}{\zeta^2} + \frac{\dt \left( \sint \xict
              \right)}{\zeta \sint} + \frac{\dphi \xicp}{\zeta \sint} \right].
\end{equation}
Euler's equation takes on the following form:
\begin{eqnarray}
\label{eq:spheroidal_Euler1}
0 &=& \vlp^2 \left[ \frac{\zeta^2 \rz \xicz}{r^2} 
        +\frac{\zeta \rt \xict }{r^2}\right]
        + 2i\vlp \frac{\Omega\zeta\sint}{r}\xicp
        - s \left(\d_s \Omega^2\right) \rz \sint
          \left[ \frac{\zeta^2\sint}{r^2} \xicz
        + \frac{\zeta\left(\rt\sint+r\cost\right)}{r^2\rz} \xict \right] \nonumber \\
  & &   - \frac{P_0}{\rho_0}\dz \left(\frac{\delta p}{P_0}\right)
        + \frac{\dz P_0}{\rho_0} \left(\frac{\delta\rho}{\rho_0} - \frac{\delta p}{P_0}\right)
        - \dz \Psi  + \left(\dz \Gz\right) \xicz + \Gz \dz \xicz + \left(\dz \Gt\right) \xict + \Gt \dz \xict, \\
\label{eq:spheroidal_Euler2}
0 &=& \vlp^2 \left[ \frac{\zeta^2 \rt \xicz}{r^2} + 
          \frac{\zeta(r^2+\rt^2) \xict}{r^2\rz} \right]
         +2i\vlp\frac{\Omega\zeta\left(\rt\sint+r\cost\right)}{r\rz}\xicp \nonumber \\
  & &    -s \left(\d_s \Omega^2\right) \left(\rt\sint+r\cost\right)
          \left[ \frac{\zeta^2\sint}{r^2}\xicz +
          \frac{\zeta\left(\rt\sint+r\cost\right)}{r^2\rz}\xict \right] \nonumber \\
  & &    -\frac{P_0}{\rho_0}\dt \left(\frac{\delta p}{P_0}\right)
         +\frac{\dt P_0}{\rho_0} \left(\frac{\delta\rho}{\rho_0} - \frac{\delta p}{P_0}\right)
         -\dt \Psi  + \left(\dt \Gz\right) \xicz + \Gz \dt \xicz + \left(\dt \Gt\right) \xict + \Gt \dt \xict, \\
\label{eq:spheroidal_Euler3}
0 &=& \vlp^2 \frac{\zeta}{\rz} \xicp
         -2i\vlp\frac{\Omega\zeta^2\sint}{r}\xicz
         -2i\vlp\frac{\Omega\zeta\left(\rt\sint+r\cost\right)}{r\rz}\xict \nonumber \\
  & &    -\frac{P_0}{\rho_0}\frac{\dphi}{\sint} \left(\frac{\delta p}{P_0}\right)
         -\frac{\dphi \Psi}{\sint}
         + \Gz \frac{\dphi\xicz}{\sint} + \Gt \frac{\dphi\xict}{\sint},
\end{eqnarray}
where $s=r\sint$ is the distance to the rotation axis. Poisson's equation becomes:
\begin{equation}
\label{eq:spheroidal.Poisson_all_lagrange}
0  =   \frac{r^2 + \rt^2}{r^2 \rz^2}  \dzz \Psi
      + \cz \dz \Psi 
      - \frac{2\rt}{r^2 \rz} \dzt \Psi
      + \frac{1}{r^2}  \lapl_{\theta \phi} \Psi 
      - 4\pi \left\{\rho_0\frac{\delta\rho}{\rho_0} 
      - \frac{\zeta^2}{r^2\rz}\left(\dz\rho_0 \xicz 
      + \frac{\dt\rho_0\xict}{\zeta}\right)\right\},
\end{equation}
where
\begin{eqnarray}
\cz &=& \frac{1}{r^2 \rz^3} \left( 2 \rz \rt \rzt - r^2 \rzz - \rz^2 \rtt 
+ 2 r\rz^2 - \rt^2 \rzz -  \rz^2 \rt \cott \right), \\
\lapl_{\theta\phi} &=& \dtt + \cott \dt + \frac{1}{\sin^2 \theta}\dpp.
\end{eqnarray}
As was pointed out in Sect.~\ref{sect:pulsation_equations}, the relative
Lagrangian density perturbation, $\delta \rho/\rho_0$, can be eliminated in
favour of the relative Lagrangian pressure perturbation, $\delta p/P_0$,
thanks to the adiabatic relation, Eq.~(\ref{eq:adiabatic}).

\subsection{Mechanical boundary condition}
\label{sect:mechanical_bc}

As explained in Sect.~\ref{sect:boundary_conditions}, the mechanical
boundary condition is obtained by calculating the dot product between
$\vect{E}^{\zeta}$ and Euler's equation, and cancelling out the vertical
gradient of $\delta p/P_0$. Furthermore, the quantity $\dz \xicz$ is eliminated
through the continuity equation, and the terms $\Gt$ and $\dt\Gt$ vanish at the
surface. In spheroidal components, one obtains:
\begin{eqnarray}
0 &=& \vlp^2 \frac{\zeta^2}{r^2\rz} \xicz
   +  2i\vlp\Omega\frac{\zeta\left(r\sint-\rt\cost\right)}{r^2\rz^2}\xicp
   - s\ds\left(\Omega^2\right)\frac{r\sint-\rt\cost}{r\rz}
      \left(\frac{\zeta^2\sint}{r^2}\xicz
   +  \frac{\zeta\left(\rt\sint+r\cost\right)}{r^2\rz}\xict\right)  \nonumber \\
  & &+\left(\frac{r^2+\rt^2}{r^2\rz^2} \frac{\dz P_0}{\rho_0}
   -  \frac{\rt}{r^2\rz} \frac{\dt P_0}{\rho_0}\right)
      \left(\frac{\delta \rho}{\rho_0} - \frac{\delta p}{P_0}\right)
   -  \frac{r^2+\rt^2}{r^2\rz^2} \dz \Psi
   +  \frac{\rt}{r^2\rz} \dt \Psi
   -  \frac{r^2+\rt^2}{\zeta^2\rz}\Gz\frac{\delta \rho}{\rho_0}
   +  \frac{r^2+\rt^2}{r^2\rz^2}\left(\dz\Gz - \frac{2\Gz}{\zeta}\right) \xicz \nonumber \\
  & &-\frac{\rt}{r^2\rz} \left[\left(\dt \Gz\right)\xicz + \Gz \dt \xicz\right]
   +  \frac{r^2+\rt^2}{r^2\rz^2} \left[ \left(\dz\Gt\right)\xict 
   -  \Gz \frac{\dt \left(\sint\xict\right)}{\zeta\sint}
   -  \Gz \frac{\dphi\xicp}{\zeta\sint}\right]
\end{eqnarray}

\section{Lagrangian perturbation to the effective gravity}
\label{sect:dgeff_appendix}

As was explained in Sect.~\ref{sect:dgeff}, the Lagrangian perturbation to the
effective gravity, $\delta\geff$, is deduced from the vectorial Lagrangian
perturbation to the effective gravity, $\delta\vgeff$, via the relation $\delta
\geff = -\n \cdot \delta \vgeff$, where $\n$ is the outward normal at the
surface.  Furthermore, $\delta\vgeff$ includes the Lagrangian perturbation to
the gradient of the gravitational potential and the acceleration of a particle
tied to the surface, resulting from the oscillatory motions.  After adding and
subtracting $\vect{\xi} \cdot \grad \left( s\Omega^2 \es \right)$ in order to
introduce the equilibrium effective gravity, a vectorial expression is obtained
in Eq.~(\ref{eq:dvgeff}), and is reproduced here for convenience:
\begin{equation}
\delta \vgeff = - \grad \Psi + \vect{\xi} \cdot \grad \vgeff
                + (\omega+m\Omega)^2 \vect{\xi} 
                - 2i(\omega+m\Omega) \vect{\Omega}\times \vect{\xi}
                - \vect{\Omega} \times \left(\vect{\Omega} \times \vect{\xi} \right)
                - \vect{\xi} \cdot \grad \left(s\Omega^2\es\right)
\label{eq:dvgeff_appendix}
\end{equation}
In what follows, we will go through the above equation one term at a time in
order to obtain explicit expressions for the dot product of $\n$ with each one.

The Eulerian perturbation to gravity is obtained through tensor analysis:
\begin{equation}
-\n \cdot \grad \Psi = -\frac{r\rz \vect{E}^{\zeta} \cdot \left(\d_i \Psi\right)
         \vect{E}^i}{\left(r^2+\rt^2\right)^{1/2}}
         = -\frac{r\rz g^{\zeta i} \d_i \Psi}{\left(r^2+\rt^2\right)^{1/2}}
         = -\frac{\left(r^2+\rt^2\right)^{1/2}}{r\rz} \dz \Psi
         + \frac{\rt}{r\left(r^2+\rt^2\right)^{1/2}} \dt \Psi
\end{equation}
where we have used the relation $\n = \frac{r\rz}{\left(r^2 +
\rt^2\right)^{1/2}}\vect{E}^{\zeta}$.  Furthermore, we have used Einstein's
summation convention on repeated indices.

Before dealing with the next term, it is useful to introduce the contravariant
components of the Lagrangian displacement, which we distinguish from the
components given in Eq.~(\ref{eq:alternate_components}) by placing a tilde over
the top:
\begin{equation}
\vect{\xi} = \txicz \Ez + \txict \Et + \txicp \Ep.
\end{equation}
where $(\Ez,\,\Et,\,\Ep)$ is given in Eq.~(\ref{eq:covariant_basis}). We also
introduce the covariant components of the effective gravity:
\begin{equation}
\vgeff = - g_{\zeta}^{\mathrm{eff}} \vect{E}^{\zeta} - g_{\theta}^{\mathrm{eff}} \vect{E}^{\theta}.
\end{equation}
From Eq.~(\ref{eq:geff}), it is straightforward to see that $g_i^{\mathrm{eff}}
= -\partial_i P_0/\rho_0$.  Furthermore, $g_{\zeta}^{\mathrm{eff}}=\frac{r\rz}
{\left(r^2+\rt^2\right)^{1/2}} \geff$ and $g_{\theta}^{\mathrm{eff}} \equiv 0$
at the stellar surface.

In tensorial notation, the term $\n \cdot \left\{ \vect{\xi} \cdot \grad \vgeff
\right\}$ becomes:
\begin{equation}
\n \cdot \left\{ \vect{\xi} \cdot \grad \vgeff \right\} = 
\frac{r\rz}{\left(r^2 + \rt^2\right)^{1/2}}
\vect{E}^{\zeta} \cdot \left\{ \vect{\xi} \cdot \grad \vgeff \right\} = 
-\frac{r\rz}{\left(r^2 + \rt^2\right)^{1/2}}
\tilde{\xi}^i \left( \d_i g_j^{\mathrm{eff}} - \Gamma_{ij}^k g_k^{\mathrm{eff}} \right) g^{\zeta j},
\label{eq:n_dot_xi_grad_vgeff_preliminary}
\end{equation}
where $g^{ij} = \vect{E}^i \cdot \vect{E}^j$ denotes the contravariant
components of the metric tensor, and $\Gamma_{ij}^k = \left(\d_i
\vect{E}_j\right) \cdot \vect{E}^k = - \left(\d_i \vect{E}^k\right) \cdot
\vect{E}_j$ the Christoffel coefficients. 
Equation~(\ref{eq:n_dot_xi_grad_vgeff_preliminary}) can be simplified if we use
the following relation:
\begin{equation}
\div \vgeff = -g^{ij} \left( \d_i g_j^{\mathrm{eff}} 
            - \Gamma_{ij}^k g_k^{\mathrm{eff}} \right)
            = -4\pi G\rho_0 + \div \left(s\Omega^2\es\right),
\end{equation}
which is a modified form of Poisson's equation. The result is:
\begin{equation}
\n \cdot \left\{ \vect{\xi} \cdot \grad \vgeff \right\}
     = \frac{r\rz}{\left(r^2 + \rt^2\right)^{1/2}} \left\{
     \txicz \left[ -4\pi G \rho_0 + \div \left( s\Omega^2\es \right) \right]
     + \txicz \left( g^{\zeta\theta} \dt g_{\zeta}^{\mathrm{eff}}
     - A g_{\zeta}^{\mathrm{eff}} \right)
     + \txict \left( -g^{\zeta\zeta} \dt g_{\zeta}^{\mathrm{eff}}
     + B g_{\zeta}^{\mathrm{eff}} \right)\right\},
\end{equation}
where
\begin{eqnarray}
A &=& g^{\zeta\theta} \Gamma_{\zeta\theta}^{\zeta}
   + g^{\theta\theta} \Gamma_{\theta\theta}^{\zeta}
   + g^{\phi\phi} \Gamma_{\phi\phi}^{\zeta}
   = \frac{-2r^2\rz-\rz\rt^2+r\rz\rtt-r\rt\rzt+r\rz\rt\cott}{r^3\rz^2} \\
B &=& g^{\zeta\zeta} \Gamma_{\zeta\theta}^{\zeta}
   + g^{\zeta\theta} \Gamma_{\theta\theta}^{\zeta}
   = \frac{\rz\rt^3+r^3\rzt+r\rt^2\rzt-r\rz\rt\rtt}{r^3\rz^3},
\end{eqnarray}
and where we have made use of the following simplifications: $g_{\theta}^{\mathrm{eff}} =
\dt g_{\theta}^{\mathrm{eff}} = 0$ on the surface, $g^{ij} = g^{ji}$ and
$\Gamma_{ij}^k = \Gamma_{ji}^k$. The above expression can then be re-expressed in
terms of $\xir$, $\xit$ and $\geff$ to yield:
\begin{eqnarray}
\n \cdot\left\{ \vect{\xi} \cdot \grad \vgeff \right\} &=&
       \frac{r\xir - \rt\xit}{\left(r^2+\rt^2\right)^{1/2}}
       \left\{ -4\pi G \rho_0 + \div \left( s\Omega^2\es \right) \right\}
    +  \xir \left\{ -\frac{\rt\dt\geff}{r^2+\rt^2}
    +  \frac{\left(2r-\rt\cott\right)\left(r^2+\rt^2\right)+r\rt^2-r^2\rtt}
      {\left(r^2+\rt^2\right)^2}\geff\right\} \nonumber \\
   &+& \xit \left\{-\frac{r\dt\geff}{r^2+\rt^2}
    +  \frac{r\rt\left(-2r^2-3\rt^2+r\rtt\right)+\left(r^2+\rt^2\right)\rt^2\cott}
      {r\left(r^2+\rt^2\right)^2}\geff\right\}.
\end{eqnarray}

The term $\div \left( s\Omega^2\es \right)$ takes on the following expression
for a general rotation profile, $\Omega\equiv\Omega(\zeta,\theta)$:
\begin{equation}
\div \left( s\Omega^2\es \right) = 2\Omega^2 
   + \frac{\sint\left(r\sint-\rt\cost\right)}{\rz} \dz \left(\Omega^2\right)
   + \sint\cost\dt\left(\Omega^2\right).
\end{equation}
For a cylindrical rotation profile, $\Omega\equiv\Omega(s)$, it becomes:
\begin{equation}
\div \left( s\Omega^2\es \right) = 2\Omega^2 + s\d_s \left(\Omega^2\right).
\end{equation}

The term $(\omega+m\Omega)^2 \n\cdot\vect{\xi}$ is given by:
\begin{equation}
(\omega+m\Omega)^2 \n\cdot\vect{\xi} = (\omega+m\Omega)^2 \frac{r\xir-\rt\xit}
                                       {\left(r^2+\rt^2\right)^{1/2}}.
\end{equation}

The Coriolis term is given by:
\begin{equation}
-2i(\omega+m\Omega)\n\cdot\left(\vect{\Omega}\times\vect{\xi}\right) = 
  2i(\omega+m\Omega)\Omega\frac{\left(r\sint-\rt\cost\right)\xip}{\left(r^2+\rt^2\right)^{1/2}}.
\end{equation}

The last two terms are more conveniently treated together.  They take on the
following expression for a general rotation profile:
\begin{equation}
- \vect{\Omega} \times \left( \vect{\Omega} \times \vect{\xi} \right)
- \vect{\xi} \cdot \grad \left(s\Omega^2\es\right) =
- \left[\frac{r\xir-\rt\xit}{\rz}\sint\dz\left(\Omega^2\right)
+ \xit\sint\dt\left(\Omega^2\right)\right]\es.
\end{equation}
If the rotation profile is cylindrical, they become:
\begin{equation}
- \vect{\Omega} \times \left( \vect{\Omega} \times \vect{\xi} \right)
- \vect{\xi} \cdot \grad \left(s\Omega^2\es\right) =
-\left(\sint\xir+\cost\xit\right)s\d_s\left(\Omega^2\right)\es.
\end{equation}

Combining all of these equations together, and remembering the minus sign,
yields Eq.~(\ref{eq:dgeff}).

\section{Cancelling of simplified disk-integration factors}
\label{sect:cancelling}

Given the simplified form of the disk-integration factors given in
Eq.~(\ref{eq:visibility_francois}) (see
Sect.~\ref{sect:comparision_Francois_me}), it turns out that some of these
cancel out regardless of inclination.  To see this, one needs to start with an
explicit form for Eq.~(\ref{eq:visibility_francois}):
\begin{equation}
\label{eq:disk_integration_explicit}
D(i)\cos(\omega t + \psi) = \frac{ \Re \left\{
     \int_{\theta=0}^{\pi} \int_{\phi=-f(\theta)}^{f(\theta)} \delta \hat{T} (\theta)
     e^{i m \phi + i \omega t} \left[ r \left( \sin i \sint \cosp + \cos i \cost \right)
     - \rt \left( \sin i \cost \cosp - \cos i \sint \right)\right]
     r \sint \mathrm{d}\theta \mathrm{d}\phi \right\}}{\pi \Req^2 \left< \delta T \right>},
\end{equation}
where $\psi$ is a suitably chosen phase, $\left< \delta T \right>$ is given in
Eq.~(\ref{eq:visibility_francois}), and  $f(\theta)$ corresponds to the
visibility curve (\ie\ the border between the visible and hidden side of the
star).  The function $f$ obeys the following symmetry: $f(\theta) + f(\pi -
\theta) = \pi$.  We have made use of Eqs.~(\ref{eq:eo}) and~(\ref{eq:dS}) in
obtaining an explicit expression for $\eo \cdot \vect{\mathrm{d}S}$.  The integral in
Eq.~(\ref{eq:disk_integration_explicit}) is then split into two equal halves and
the second half is modified according to the variable changes $\theta' = \pi -
\theta$ and $\phi' = \phi - \pi$:
\begin{eqnarray}
\label{eq:disk_integration_explicit_bis}
D(i)\cos(\omega t + \psi) &=&  \Re \left\{
     \int_{\theta=0}^{\pi} \int_{\phi=-f(\theta)}^{f(\theta)} \delta \hat{T} (\theta)
     e^{i m \phi + i \omega t}  \left[ r \left( \sin i \sint \cosp + \cos i \cost \right)
     - \rt \left( \sin i \cost \cosp - \cos i \sint \right)\right]
     r \sint \mathrm{d}\theta \mathrm{d}\phi \right. \nonumber \\ & & 
     + \int_{\theta'=0}^{\pi} \int_{\phi'=f(\theta')-2\pi}^{-f(\theta')} (-1)^{m+1} \delta \hat{T} (\pi - \theta')
     e^{i m \phi' + i \omega t} \left[r \left( \sin i \sint' \cosp' + \cos i \cost' \right)
      \right.\nonumber \\ & & \left. \left.
     - \,r_{\theta'} \left(\sin i \cost' \cosp' - \cos i \sint' \right) \right]
     r \sint' \mathrm{d}\theta' \mathrm{d}\phi' {\color{white}\int} \hspace*{-3mm} \right\}/
     \left(2\pi \Req^2 \left< \delta T \right>\right),
\end{eqnarray}
where we have made use of the relation $f(\theta) + f(\pi-\theta) = \pi$.  The two halves can be
combined to give a single integral over the \textit{entire} stellar surface only
if $\delta \hat{T} (\theta) = (-1)^{m+1} \delta \hat{T} (\pi - \theta)$.  If we
assume this is the case, we can then see under conditions the integral vanishes.
We specifically look at the integration over $\phi$, which now is between the
bounds $0$ and $2\pi$.  Remembering that $\int_0^{2\pi} e^{im\phi} \cosp \mathrm{d}\phi =
0$ if $|m| \neq 1$ and that $\int_0^{2\pi} e^{im\phi} \mathrm{d}\phi = 0$ if $m \neq 0$,
we deduce the second condition for cancelling the disk-integration factor, \ie\
$|m| \geq 2$.

\section{Normalisation of multi-colour visibilities}
\label{app:normalisation}

In order to find a normalisation which minimises the distances between a set of
multi-colour visibilities, we start with the cost function given in
Eq.~(\ref{eq:cost_simple}) and include an additional term so as to enforce the
constraint $\sum_{i=1}^{b} W_i^2 = 1$.  Without loss of generality, we work with
the normalised components, $\tilde{V}_i^j$, instead of the original ones:
\begin{equation}
J = \sum_{j=1}^{N} \sum_{i=1}^{b} \left( \tilde{A}_j \tilde{V}_i^j - W_i \right)^2
  + \Lambda\left(1 - \sum_{i=1}^{b} W_i^2 \right),
\label{eq:cost}
\end{equation}
where $\Lambda$ represents a Lagrange multiplier. Setting the derivatives,
$\partial J/\partial \tilde{A}_j$, $\partial J/\partial W_i$, and $\partial
J/\partial \Lambda$, to zero leads to the following system:
\begin{eqnarray}
\label{eq:Aj_app}
\tilde{A}_j &=& \frac{\tilde{\vect{V}}^j \cdot \vect{W}}
                   {\tilde{\vect{V}}^j \cdot \tilde{\vect{V}}^j}, \\
\label{eq:W}
(N-\Lambda) \vect{W} &=& \sum_{j=1}^{N} \tilde{A}_j \tilde{\vect{V}}^j, \\
\vect{W} \cdot \vect{W} &=& 1,
\end{eqnarray}
where we've used vectorial notation for conciseness and where $\vect{A} \cdot
\vect{B}$ represents the dot product $\sum_{i=1}^b A_i B_i$. Using the
normalised components $\tilde{V}_i^j$ allows us to simplify
Eq.~(\ref{eq:Aj_app}) to $\tilde{A}_j = \tilde{\vect{V}}^j \cdot \vect{W}$. 
When combined with Eq.~(\ref{eq:W}), this yields:
\begin{equation}
(N-\Lambda) \vect{W} = \sum_{j=1}^{N} \left(\tilde{\vect{V}}^j\cdot\vect{W}\right)\tilde{\vect{V}}^j.
\end{equation}
This last equation is in fact an eigenvalue problem where $N-\Lambda$ is the
eigenvalue and $\vect{W}$ the eigenvector. In order to determine which
eigensolution yields the minimal value for $J$, we develop the cost function as
follows:
\begin{equation}
J = \sum_{j=1}^N \left\{ \tilde{A}_j^2 \tilde{\vect{V}}^j \cdot \tilde{\vect{V}}^j
    - 2 \tilde{A}_j \tilde{\vect{V}}^j \cdot \vect{W} + \vect{W} \cdot \vect{W}\right\}
  = \sum_{j=1}^N \left\{1 - \tilde{A}_j^2\right\} = N - \sum_{j=1}^N \tilde{A}_j^2,
\end{equation}
where we've used the simplifications $\tilde{\vect{V}}^j \cdot
\tilde{\vect{V}}^j = 1$, $\tilde{A}_j = \tilde{\vect{V}}^j \cdot \vect{W}$, and
$\vect{W} \cdot \vect{W} = 1$.  In order to simplify the term $\sum_{j=1}^N \tilde{A}_j^2$,
we calculate the dot product between $\vect{W}$ and Eq.~(\ref{eq:W}):
\begin{equation}
(N-\Lambda) \vect{W} \cdot \vect{W} = N - \Lambda
      = \sum_{j=1}^N \tilde{A}_j \left( \tilde{\vect{V}}^j \cdot \vect{W} \right)
      = \sum_{j=1}^N \tilde{A}_j^2.
\end{equation}
Hence,
\begin{equation}
J = N - (N - \Lambda) = \Lambda.
\end{equation}
Therefore, the minimal value of $\Lambda$ (and hence the maximal value of
$N-\Lambda$) corresponds to the minimal value of $J$.  The vector $\vect{W}$ is
therefore the principal component of the vector set
$\left(\tilde{\vect{V}}^j\right)_{j=1 \dots N}$ and can be found either via a
singular value decomposition of the associated matrix, or more simply through a
power iteration.
\end{document}